\documentclass{aastex61}
\usepackage[flushleft]{threeparttable}
\usepackage{color}
\usepackage{float}
\usepackage[caption = false]{subfig}
\usepackage{amsmath}
\usepackage{graphicx}
\usepackage{mathtools} 

\newcommand{\NII}{[\ion{N}{2}]}  
\newcommand{\OIII}{[\ion{O}{3}]}   

\newcommand{\Hb}{\ensuremath{{\rm H}\beta}}
\newcommand{\Ha}{\ensuremath{{\rm H}\alpha}}

\newcommand{\beq}{\begin{equation}}
\newcommand{\eeq}{\end{equation}}

\shorttitle{All-Sky Optical AGN Catalog}
\shortauthors{Zaw et al.}

\begin{document}


\title{A Uniformly Selected, All-Sky, Optical AGN Catalog}

\correspondingauthor{Ingyin Zaw}
\email{iz6@nyu.edu}

\author{Ingyin Zaw}
\affil{New York University Abu Dhabi,
    P.O. Box 129188, Abu Dhabi, United Arab Emirates}
\author{Yan-Ping Chen}
\affiliation{New York University Abu Dhabi,
    P.O. Box 129188, Abu Dhabi, United Arab Emirates}

\author{Glennys R Farrar}
\affiliation{Center for Cosmology and Particle Physics, Physics Department, New York University,
        New York, NY 10003}







\begin{abstract}

We have constructed an all-sky catalog of optical AGNs with $z < 0.09$, based on optical spectroscopy, from the parent sample of galaxies in the 2MASS Redshift Survey (2MRS), a near-complete census of the nearby universe. Our catalog consists of 1929 broad line AGNs, and 6562 narrow line AGNs which satisfy the \citet{Kauffmann03} criteria, of which 3607 also satisfy the \citet{Kewley01} criteria. We also report emission line widths, fluxes, flux errors, and signal-to-noise ratios of all the galaxies in our spectroscopic sample, allowing users to customize the selection criteria. Although we uniformly processed the spectra of galaxies from a homogeneous parent sample, inhomogeneities persist due to the differences in the quality of the obtained spectra, taken with different instruments, and the unavailability of spectra for $\sim$20\% of the galaxies. We quantify how the differences in spectral quality affect not only the AGN detection rates but also broad line to narrow line AGN ratios. We find that the inhomogeneities primarily stem from the continuum signal-to-noise (S/N) in the spectra near the emission lines of interest. We fit for the AGN fraction as a function of continuum S/N and assign AGN likelihoods to galaxies which were not identified as AGNs using the available spectra. This correction results in a catalog suitable for statistical studies. This work also paves the way for a truly homogeneous and complete nearby AGN catalog by identifying galaxies whose AGN status needs to be verified with higher quality spectra, quantifying the spectral quality necessary to do so.

\end{abstract}




\section{Introduction} \label{sec:intro}

Optical line analysis is currently the most robust method available for identifying active galactic nuclei (AGNs) \citep{Padovani17}. The presence or absence of broad Balmer lines is also the best indicator of the orientation (face-on vs. edge-on) of the accretion disk and parsec-scale dusty ``torus". Uniformly selected all-sky AGN catalogs at other wavelengths, such as the hard X-ray {\it Swift}-BAT AGN catalog \citep{BATcat}, the infrared {\it WISE} AGN catalog \citep{WISEcat}, the all-sky near-by radio galaxy catalog \citep{radiocat}, and even the spectroscopic follow-up of the BAT catalog \citep[e.g.,][]{Koss17}, are available. However, an all-sky uniformly selected optical AGN catalog has not previously been constructed. 

An all-sky optical AGN catalog is important for comparisons with all-sky catalogs at other wavelengths. Studies have shown that AGN identification in different wavelengths select overlapping but non-identical samples of AGNs \citep[e.g.,][]{Azadi17}. Furthermore, optical spectroscopy could be missing a large fraction of obscured AGNs detected in the infrared \citep[e.g.,][]{Goulding09}. Some studies have suggested that optical AGNs are a disjoint population from X-ray AGNs \citep{Arnold09} (but see \citep{Dai2017}, which refutes this) and there is debate as to the presence of a bimodality in radio loudness \citep[e.g.,][]{Singal13}. However, these analyses have been performed only on a small portion of the sky and/or with a small sample. A complete census of optical AGNs in the nearby universe is necessary to determine the full extent of the overlap, or lack thereof, between AGNs selected at different wavelengths. It will also provide a first step in understanding whether these differences are simply due to selection biases or more fundamental differences in the physics of AGNs, such as whether AGNs identified in different wavebands result from differences in accretion states. In addition, an all-sky optical AGN catalog is essential for statistical correlation studies with phenomena possibly linked to AGN such as ultra-high energy cosmic rays \citep[e.g.,][]{AugerSci, AugerCor} or to search for phenomena found only in some AGNs, such as water maser emission primarily found in a subset of narrow line AGNs \citep[e.g.,][]{Braatz04, Kondratko06, Zhu11}.

The most complete listing of optical AGNs at all redshifts is the Veron-Cetty \& Veron Catalogue of Quasars and Active Galactic Nuclei, 13th edition \citep[VCV catalog]{vcv}, a collection of optical AGN candidates reported in literature. While this catalog is a superset of all other optical AGN catalogs, it is incomplete and highly inhomogeneous \citep{zfb11}, since the AGN candidates are detected by different instruments from observations which target different parts of the sky with different depths. Furthermore, it is also not ``pure" and contains star forming galaxies previously misclassified as AGNs \citep[]{zfg09, tzf12}. It was used in statistical correlation studies such as comparison with the arrival direction of ultra-high energy cosmic rays (UHECRs) \citep[e.g.,][]{AugerSci, AugerCor}. Unfortunately, the shortcomings of the catalog used for the correlation studies make the tantalizing and persistent, but low significance (2- to 2.7-$\sigma$), result impossible to interpret reliably. The sources of UHECRs, with energies above $\sim 5 \times 10^{19}$ eV, have the double constraint that they must be energetic enough to accelerate particles to such energies and be within distances of order 100 Mpc due to the Greisen-Zatsepin-Kuzmin (GZK) effect. AGNs remain a leading candidate for the acceleration of UHECRs but a uniform, complete, nearby AGN catalog is needed to determine whether UHECRs originate from AGNs.

We have constructed an all-sky, uniformly-selected and processed AGN catalog from optical spectra of the galaxies in the 2MASS Redshift Survey \citep[2MRS]{2MRS}, a near-complete map of the nearby (out to z$\lesssim$0.09) universe and a uniform parent sample. We have collected spectra for $\sim$80\% of the galaxies and after initial quality cuts, apply uniform spectral fitting and AGN identification procedures to the remaining sample. We also report emission line widths, fluxes, flux errors, and signal-to-noise ratios, allowing users to customize the selection criteria. Although our parent galaxy sample is complete, the spectral sample was compiled from different sources. The intrinsic differences in spectral qualities of our subsamples present a challenge to constructing a homogeneous catalog. We take advantage of this diversity to quantify the effects of data quality on AGN identification. This is important not only for the current analysis but in comparing AGN detection rates and broad line to narrow line AGN ratios obtained with different optical surveys. Furthermore, we develop a method to statistically compensate for the inhomogeneity and incompleteness in AGN identification due to low spectral signal-to-noise.

This paper is organized as follows. Section~\ref{sec:sample} describes our spectroscopic sample. Section~\ref{sec:AGNID} describes the methods we used to isolate the emission lines and identify AGNs. The validation of our method and cross checks are presented in Section~\ref{sec:validation}. Section~\ref{sec:catalog} describes our AGN catalog. A study of the impact of spectral signal-to-noise and resolution on AGN detection rates is presented in Section~\ref{sec:detectionrates} with a method to statistically correct for the resulting incompleteness and inhomogeneity presented in Section~\ref{sec:statcat}. Our findings are summarized in Section~\ref{sec:conclusions}.

\section{Spectroscopic Sample of Galaxies}
\label{sec:sample}

\begin{figure}
\includegraphics[scale=0.75]{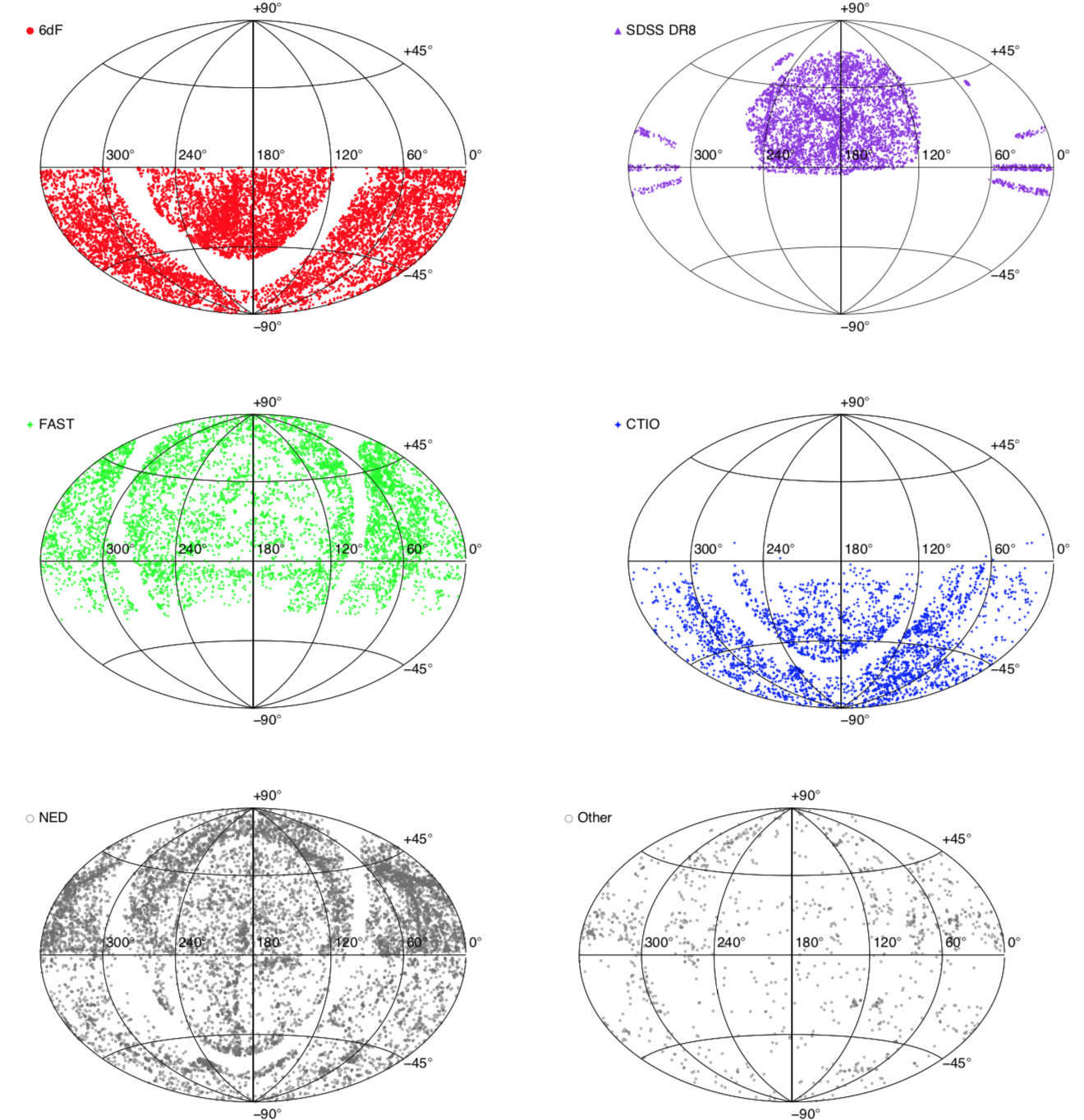}
\caption{The sky distributions of the 2MRS subsamples: 6dF (top, left), SDSS (top, right), FAST (middle, left), CTIO (middle, right), NED galaxies (bottom, left), and galaxies from other sources (bottom, right).}
\label{fig:subsamples}
\end{figure}

We take the 2MASS Redshift Survey \citep[2MRS]{2MRS} as the parent sample for constructing our all-sky optical AGN catalog. The 2MRS galaxy catalog, a compilation of redshifts for 43,533 galaxies with $K_s \leq 11.75$ (mean $z = 0.03$) and Galactic latitude $|b| > 5^\circ$ ($8^\circ$ toward the Galactic bulge) in the Two Micron All-Sky Survey \citep[2MASS]{2MASS}, is a near-complete (91\% of the sky) census of galaxies in the near-by ($z\lesssim0.09$) universe (above the flux limit). Nearly half of redshifts in 2MRS were taken from large surveys for which the digital spectra are publicly available, namely the 6dF Galaxy Survey \citep[6dF]{6dF04, 6dFGS} and the Sloan Digital Sky Survey \citep[SDSS]{SDSS}. For an additional quarter of galaxies for which no redshift measurements were available, the 2MRS team took spectra with the the long-slit FAst Spectrograph for the Tillinghast telescope \citep[FAST]{FAST} on the 1.5 m Tillinghast reflector at the Fred L. Whipple Observatory (FLWO) for northern galaxies and  the CHIRON spectrograph on the 1.5 m SMARTS telescope and R-C grating spectrograph on the V.M. Blanco 4.0 m telescope at the Cerro Tololo Interamerican Observatory (CTIO) in the South. For the remaining quarter of galaxies, redshifts were compiled from the NASA Extragalactic Database (NED), from more than 550 references, and the optical spectroscopic CfA Redshift Catalog (ZCAT). The sky distributions of these different 2MRS subsamples is shown in Figure~\ref{fig:subsamples}. Optical spectra are available for a majority of the 2MRS galaxy sample. 

We have collected optical spectra for $\sim$80\% of the 2MRS galaxies by gathering the public digital optical spectra from SDSS and 6dF, obtaining the private FAST and CTIO spectra (courtesy of Lucas Macri), and downloading the spectra available in the NED database. The spectra were matched to the 2MRS objects either from their identification numbers, when possible, or counterparts were identified by requiring that the angular separation and redshift difference be within the errors of the respective samples. The 3229 spectra we compiled from NED were taken with tens of different instruments and have disparate spectral coverage, resolution, and quality. Since it is virtually impossible to treat these spectra uniformly, we do not include them in our sample. Our spectroscopic sample for AGN identification has four main subsamples, 6dF, SDSS, FAST, and CTIO, in descending order by size, as listed in Table~\ref{tab:sources}.

\begin{table}
 \caption{The Parent 2MRS Sample}
 \label{tab:sources}
 \begin{center}
 \begin{tabular}{|l|r|r|r|c|c|}
 \hline
  \multicolumn{1}{|c|}{2MRS subsample} &
  \multicolumn{1}{c|}{N$_{\rm 2MRS}$} &
  \multicolumn{1}{c|}{N$_{\rm spectra}$} &
  \multicolumn{1}{c|}{N$_{\rm sample}$} &
  \multicolumn{1}{c|}{Wavelength} & 
  \multicolumn{1}{c|}{Resolution}  \\
\    &\   & \  & \  & range (\AA) & \\ 
\hline
  6dF & 11763 &11762 & 10356 & V: 4000-5500  & FWHM$_{V}$: 5-6\ \AA\\
  \     & \     &  \    & \ &R: 5500-7500 &  FWHM$_{R}$: 9-12\ \AA\\
  SDSS & 7069 & 7069 & 7069 & 3800-9200 & $\frac{\lambda}{\Delta\lambda}\sim$1800-2000\\
  FAST & 7590 &7547 & 6271 & 3500-7400& FWHM: 5\ \AA \\
  CTIO & 3293 & 3286 & 2851 & 3700-7200 & FWHM: 7.0\ \AA\\
  NED & 12694 & 3229 & -- & -- & --\\
  Others & 1124 & 0 & -- & -- & --\\
  total & 43533 & 32893 & 26547 & --& --\\
\hline
\end{tabular}
\end{center}
   \begin{tablenotes}
      \small
      \item The table lists the total number of 2MRS galaxies in each subsample, N$_{\rm 2MRS}$, the number for which we have spectra, N$_{\rm spectra}$, and the number in our final spectral sample after the Galactic plane ($|b|>10^\circ$) and telluric contamination cuts have been applied, N$_{\rm sample}$, as well as the wavelength range and resolution of each processed subsample. The 6dF spectra were taken in two bands, designated V-band and R-band, and each has a different resolution. 
\end{tablenotes}
\end{table}

\begin{figure}[htb]
\begin{center}
\includegraphics[scale=0.6,angle=90]{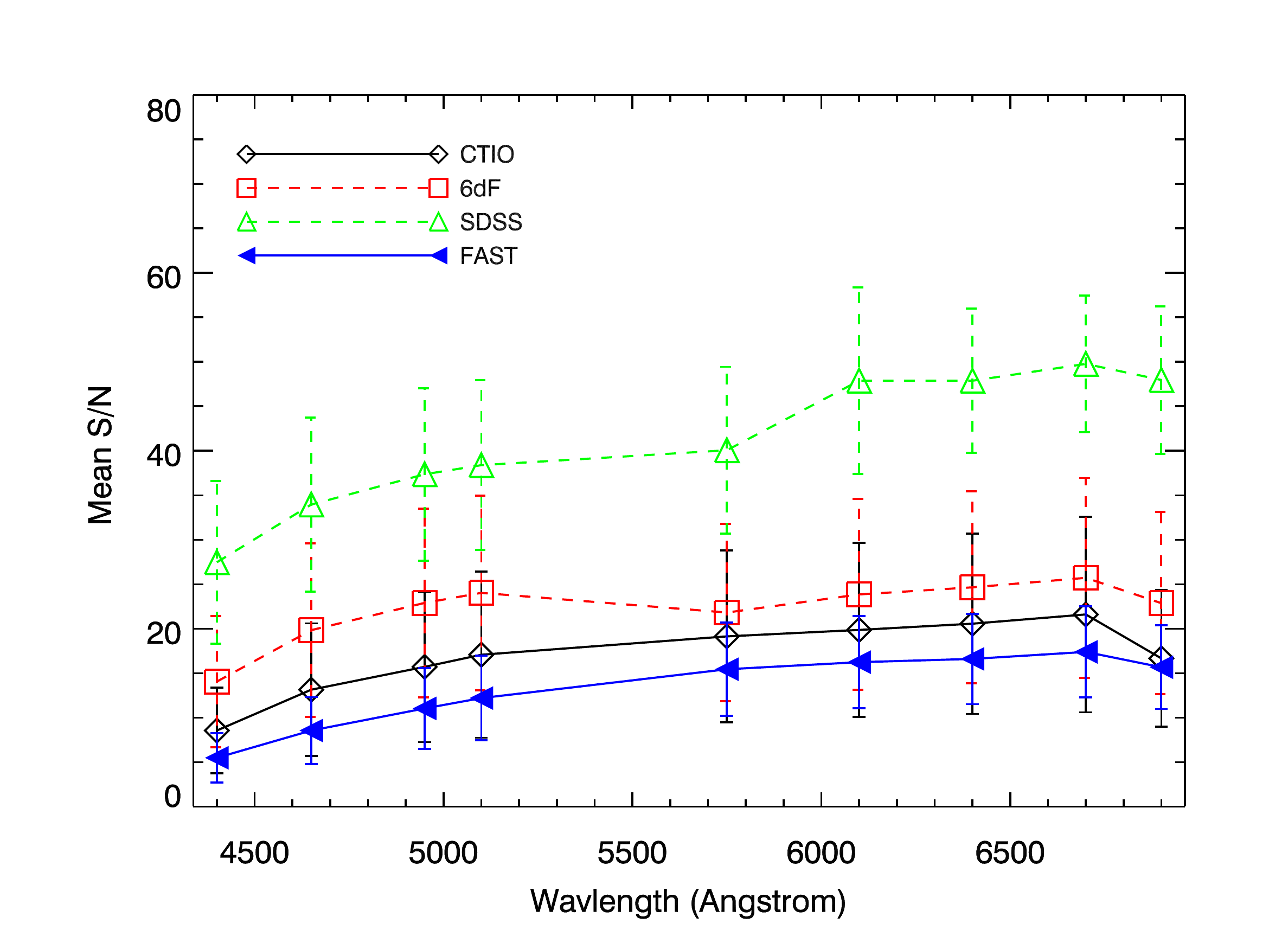}
\end{center}
\caption{Signal-to-noise ratio (S/N) of the subsamples vs. wavelength. We define the S/N in each wavelength bin as the data divided by the error provided in the error spectrum. The symbols are the mean and error bars are the standard deviation of the S/N distributions.}
\label{fig:sampleSNR}
\end{figure}

The detailed description of the differences in the spectra of our subsamples can be found in \citet{2MRS}. We have listed the the relevant parameters, namely the wavelength ranges and spectral resolution, of the spectra from each subsample (6dF, SDSS, FAST, and CTIO) in Table~\ref{tab:sources}. Each 6dF spectrum is composed of two spectra with different spectral resolution, the V-band spectrum with the wavelength range 3900-5600 \AA\ and the R-band spectrum with the wavelength range 5400--7500 \AA. In addition, the spectra in the different subsamples have different signal-to-noise (S/N) ratios. We define the S/N in each wavelength bin as the data divided by the error provided in the error spectrum. The S/N values are wavelength dependent. The mean and standard deviations of the S/N for the subsamples at various (continuum) wavelengths are shown in Figure~\ref{fig:sampleSNR}. 

Overall, the SDSS spectra have the widest wavelength coverage, the best spectral resolution, and the highest (average) S/N. The 6dF, FAST, and CTIO spectra have lower S/N than SDSS spectra but are comparable to each other. Furthermore, the SDSS subsample has absolute flux calibration while the other three subsamples do not.\\  

The telluric B-band (6860-6890 \AA), which results from the absorption of light by molecules in the Earth's atmosphere, can overlap with the \NII -\Ha\ complex in the redshift range $0.0407 < z < 0.0511$\footnote{Note: These redshifts are in the rest frame of the Earth and fit from the spectra. These redshifts change throughout the year due to the motion of the Earth around the Sun. In order to remove this time dependence, the values listed in the catalog are ``heliocentric" redshifts, i.e. in the reference frame of the Sun and constant throughout the year.} This overlap will result in incorrect \NII\ and \Ha\ fluxes. While the SDSS spectra have been corrected for this telluric contamination, the 6dF, FAST, and CTIO subsamples do not have this correction applied. Therefore, we exclude the 6dF, FAST, and CTIO galaxies ($\sim$14\%) which fall within this redshift range from our sample, unless the spectra show a clear sign of broad \Ha\ emission. 

\begin{figure}[htb]
\includegraphics[scale=0.7]{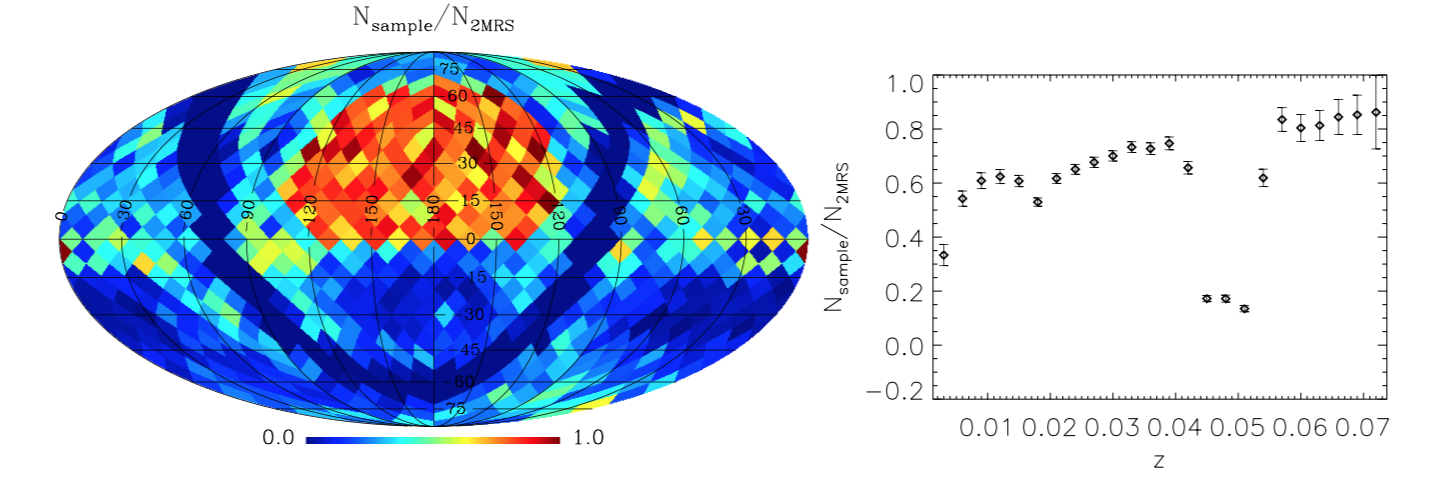}
\caption{The sky (left, divided into 768 equal area regions) and redshift (right) distributions of the 2MRS galaxies in our sample. Each pixel in the sky distribution and each bin in the redshift distribution shows the ratio of the number of galaxies in our spectroscopic sample to the total number of galaxies in 2MRS. The completeness of our sample is largely uniform in the southern hemisphere where 6dF dominates. In the North, the completeness is highest in the SDSS footprint. The completeness in redshift is largely uniform except for the telluric contamination range when we only have data from SDSS.}
\label{fig:sampdist}
\end{figure}

The 2MRS sample contains redshifts for galaxies within $5^\circ$ or $8^\circ$ (toward the Galactic bulge) of the Galactic Plane. However, they mainly come from the NED sample and there is higher incompleteness in this region. Therefore, we impose a larger Galactic Plane cut of $|b| > 10^\circ$ to improve the homogeneity of completeness in our sample. Our final spectral sample consists of 26,547 galaxies. The detailed breakdown of the 2MRS parent sample, number of galaxies with available optical spectra, and the final sample for AGN identification after the requirements described above, are given in Table~\ref{tab:sources}. Figure~\ref{fig:sampdist} shows the ratio of galaxies in our final spectroscopic sample to the number of galaxies in 2MRS, across the sky and in redshift. The completeness of our sample is largely uniform in the southern hemisphere where 6dF dominates. In the North, the completeness is highest in the SDSS footprint. The completeness in redshift is largely uniform except for the telluric contamination range when we only have data from SDSS.

\section{Spectral Fitting and AGN Identification}
\label{sec:AGNID}

We identify AGNs by first subtracting the stellar absorption and continuum emission from the host galaxy and then analyzing the emission lines. 

\subsection{Subtraction of Host Galaxy Contribution to the Spectrum}
\label{sec:ssps}

The optical spectrum for a galaxy contains a mix of contributions from the galaxy and the AGN, if present. We subtract the stellar absorption and continuum emission from the galaxy by modeling them using full spectrum fitting with a stellar population model. A stellar population model (SPM) is constructed from a stellar library, a set of empirical or theoretical spectra of stars with different temperatures and metallicities, assuming an initial mass function (IMF) of stars in a given burst of star formation, and integrating along an isochrone, a curve on the Hertzsprung-Russell diagram populated by stars of the same age and metallicity. Doing so yields a set of single stellar populations (SSPs), where each SSP is a spectrum of a stellar cluster with a given age and metallicity. Since there are many possible choices for stellar libraries, IMFs, and isochrones, many SPMs are available. The systematic differences in AGN identification due to the choice of SPM is explored in detail by \citet{CZF}.

SPMs based on theoretical stellar libraries are subject to model dependence on line profiles and opacity uncertainties. Empirical stellar libraries are not model dependent but need to a large number of observed stars that cover the parameter space as completely as possible. The MILES SPM \citep{MILES} is constructed from the MILES stellar library \citep{MILESlib}, currently the largest and most well calibrated empirical stellar library. The MILES stellar library consists of 985 stars with a wavelength range of 3525-7500 \AA\ and a spectral resolution of 2.5 \AA. In this work, we use the MILES SPM, with a unimodal IMF of slope 1.3, the Salpeter IMF \citep{Salpeter55}, and the Padova isochrone \citep{Padova00}. It contains SSPs with ages from 63 Myr to 17.78 Gyr and metallicities ranging between [M/H] = -2.32 and +0.22.

\begin{figure}[htb]
\begin{center}
\includegraphics[width=0.9\textwidth]{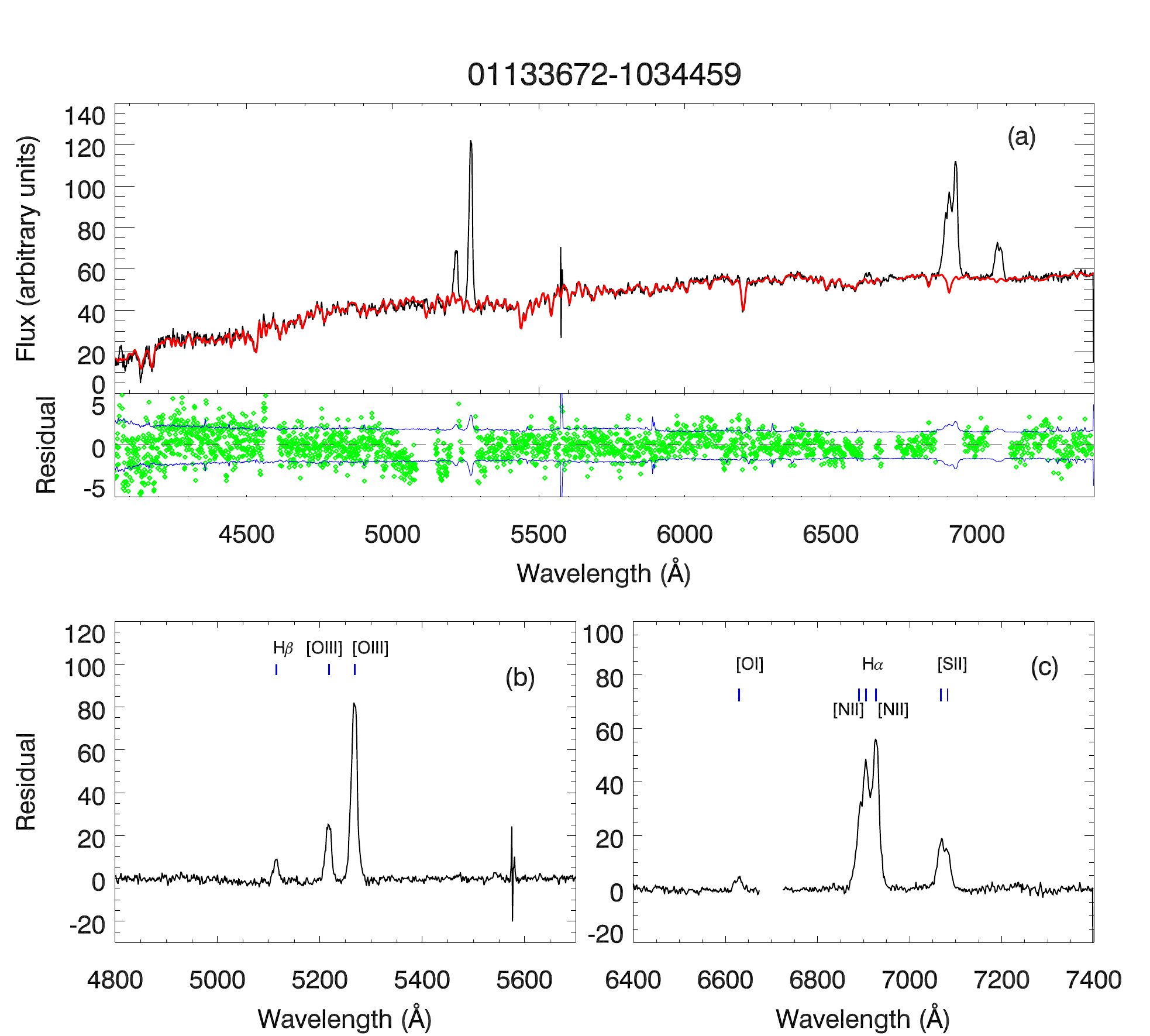}
\end{center}
\caption{Full-spectrum-fitting of an illustrative SDSS DR8 spectrum. Panel (a): DR8 data spectrum is shown as a black line, the error spectrum is shown as a blue line, and the best stellar population model fit of  host-galaxy spectrum in red. The green dots show the residual in the line-free regions. The bottom panel shows the residuals, i.e. emission lines, in the fit. Panel (b) and (c): The residual spectrum zoomed in on the two wavelength ranges with the emission lines used for AGN identification.}
\label{fig:SSPfitting}
\end{figure}

For each spectrum in our galaxy sample, wavelength ranges with emission lines and telluric contamination bands are masked. Since, each galaxy from 6dF was observed in two bands, V (4000-5500 \AA) and R (5500-7500 \AA), we combine them into a single spectrum by matching in the overlap region, and smooth the combined spectrum to the lower (R-band) resolution, before fitting. The SSPs are broadened to match the spectral resolution of the observed spectra. Each data spectrum is fit as a linear combination of the MILES SSPs using the full-spectrum-fitting program pPXF \citep{Cappellari04}. A large fraction of the spectra are fit by a linear combination of one or two SSPs and nearly all have fewer than five SSPs in the linear combination. For the SDSS spectra only an overall normalization is added to the templates. For 6dF, CTIO, and FAST spectra which do not have absolute flux calibration, the fitting algorithm adds a polynomial, of up to eighth order, to account for the instrumental effects on the spectral shape. The fitting algorithm further broadens the templates with a Gaussian kernel to mimic the stellar velocity dispersion in the integrated spectrum of the host galaxy. The fits are required to yield a physical stellar velocity dispersion, $\sigma_{\rm fit} < 1000$ km/s, in order to be considered acceptable. The fitting routine uses the error spectrum, provided with the data, to calculate a reduced $\chi^2$ value for each acceptable fit. The acceptable fit with the lowest reduced $\chi^2$ is taken to be the best fit spectrum for the galaxy. Figure~\ref{fig:SSPfitting} shows a typical spectrum from the SDSS sample (black), the error (blue), the best fit (red) and the residual spectra after subtraction in the regions around the emission lines used for AGN identification. 

For the rare cases where the template fitting fails (5 spectra in SDSS, 14 in FAST, 164 in 6dF, 386 in CTIO), we fit for the continuum in the wavelengths on either side of the emission lines of interest and interpolate. In these cases, no \Ha\ and \Hb\ stellar absorption contributions are subtracted. If the template fitting failed because there is no significant absorption, this procedure does not introduce any bias. However, if the failure is due to low spectral signal-to-noise, the \Ha\ and \Hb\ emission will be underestimated and the line ratios will be overestimated, leading to a systematically higher number of galaxies identified as AGNs. The AGNs which are identified after this ``local fitting" procedure are marked as such in the catalog and constitute only a few percent of the total AGN sample.

\newpage
\subsection{Correction of Underestimated and Missing Error Spectra}
\label{sec:missingspec}

\begin{figure}[htb]
\begin{center}
\includegraphics[width=0.45\textwidth]{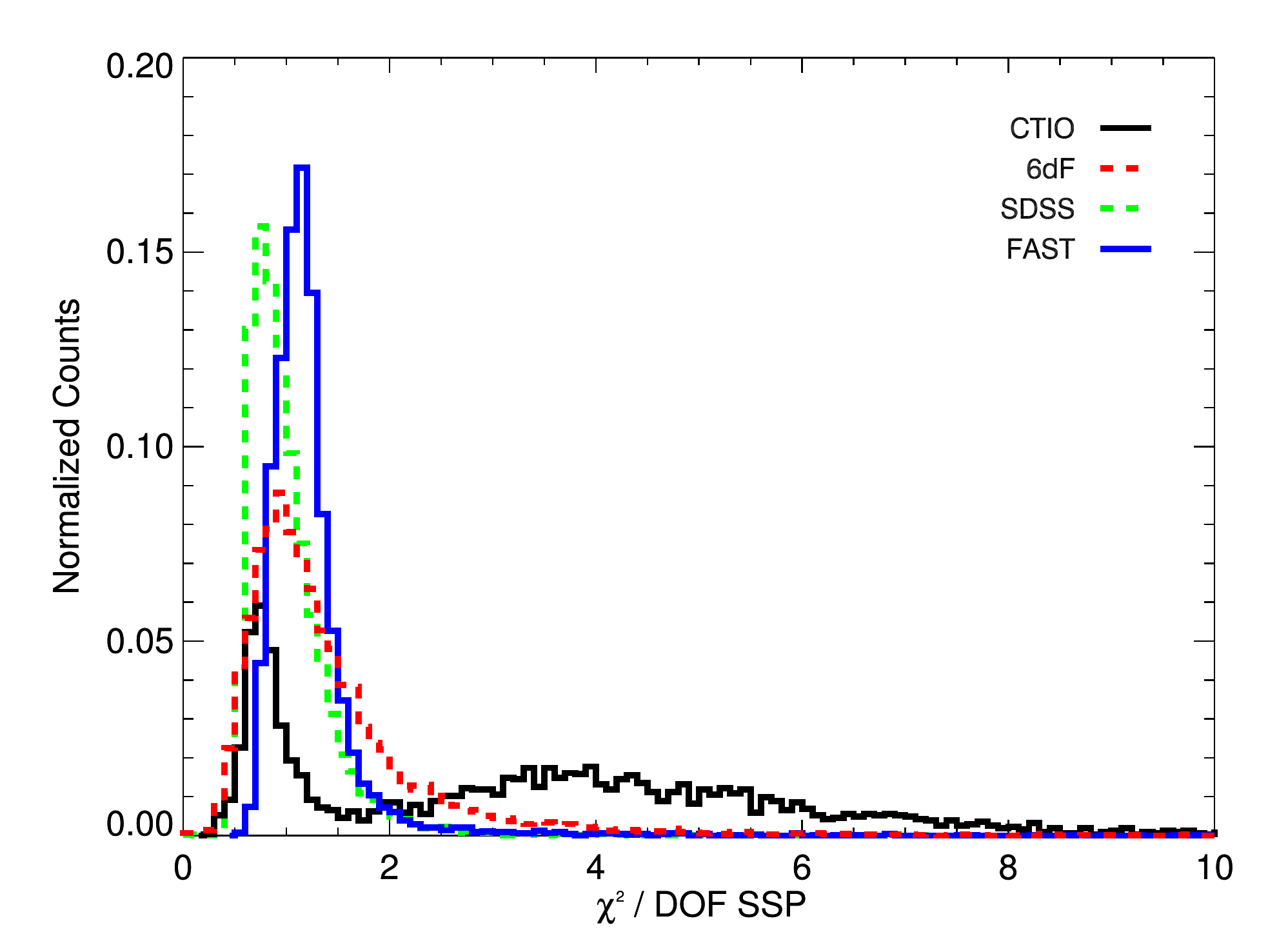}
\includegraphics[width=0.45\textwidth]{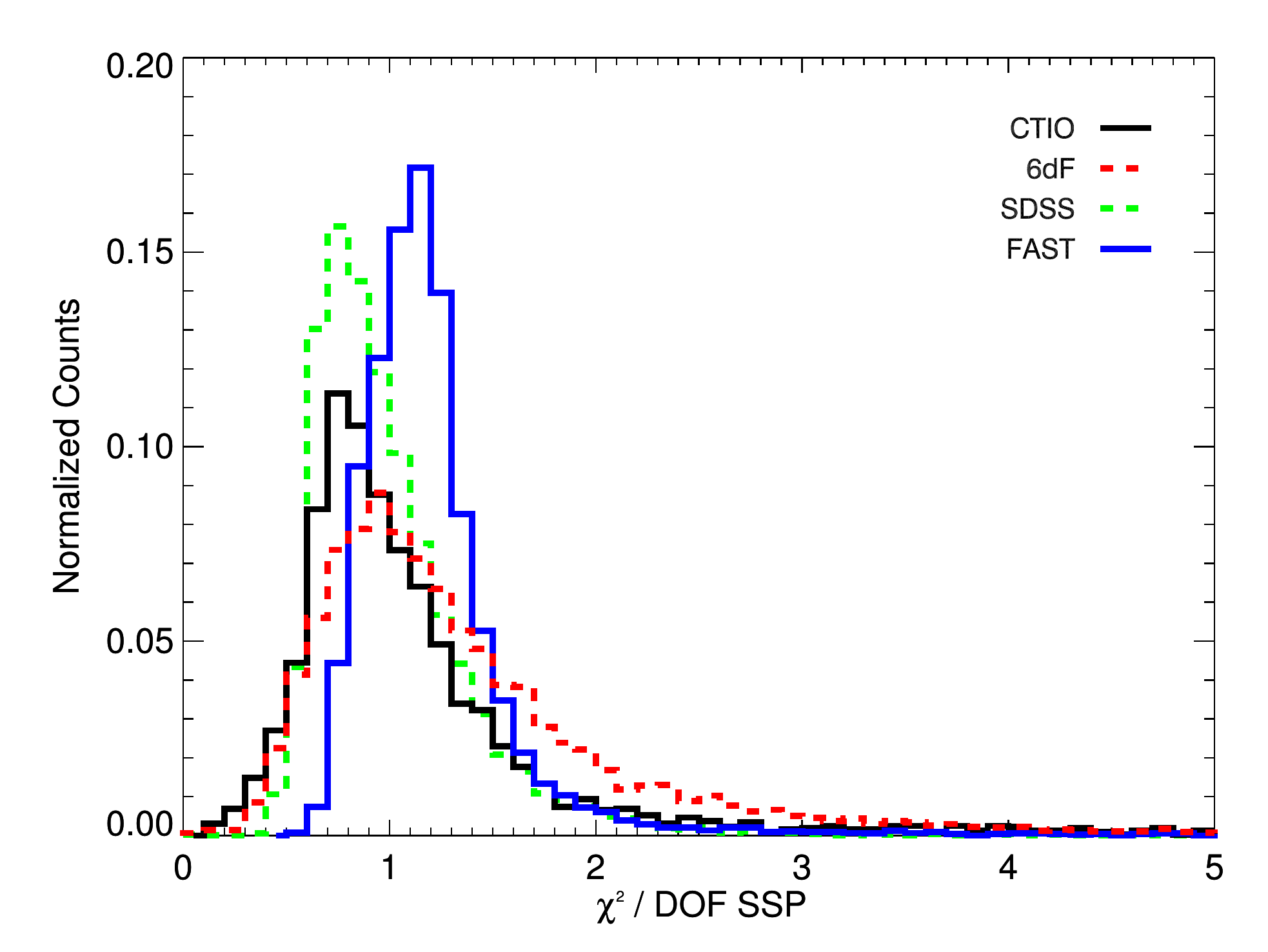}
\end{center}
\caption{Distributions of reduced $\chi^2$ from the SSP fitting before (left) and after (right) the CTIO error spectra have been fixed.}
\label{fig:SSPchi2}
\end{figure}

Figure~\ref{fig:SSPchi2} shows the distributions of reduced $\chi^2$ from the best fit SSP linear combinations. As expected, the distributions (left panel) for the SDSS, 6dF, and FAST subsamples peak near $\chi^2=1$ and follow a normal $\chi^2$ distribution. The differences in the widths are likely due to the fact that the different subsamples have different spectral resolutions, i.e. different numbers of wavelength bins, and, therefore, numbers of degrees of freedom. The CTIO subsample, however, shows a broad second peak extending to $\chi^2\sim10$. 

We examine a random selection of the CTIO spectra with large reduced $\chi^2$ values from their SSP fitting and find that the residual errors show a random (Gaussian) distribution rather than a systematic deviation from the fit. Therefore, we suspect that the errors are underestimated by a uniform multiplicative factor. Since the error spectrum is used to determine the S/N of the emission line fluxes, we need to correct the errors. 

In order to find an underlying cause for the error mis-estimation, we plot the SSP reduced $\chi^2$ values versus sky location, redshift, and date of observation. We find that on any given day or set of adjacent days, which we call an ``observing run", the reduced $\chi^2$ distribution has a single peak, but that on different observing runs, the peaks are at different values. This indicates that errors were uniformly underestimated for some of the runs. To correct for these underestimation, we fit the reduced $\chi^2$ distribution for each run with the $\chi^2$ distribution function letting the peak value vary. Then we multiply the errors for each wavelength bin of each spectrum in the run by the square root of the peak value. The right panel of Figure~\ref{fig:SSPchi2} shows the distributions of reduced $\chi^2$ after correction; the CTIO distribution has a single peak near 1.0 as expected. From the reduced $\chi^2$ distributions, we find the values that contain 99\% of the galaxies with successful fits, namely 2.55, 6.05, 3.75, and 7.15, for SDSS, 6dF, FAST, and CTIO, respectively; we keep only the galaxies with reduced $\chi^2$ less than these values.

In addition to mis-estimated errors, 218 spectra in the CTIO sample did not have associated error spectra. For those galaxies, we used the average of the corrected error spectra for the observing run of the galaxy. Similarly, 35 FAST spectra did not have error spectra. For those we use the average of the error spectra in the FAST sample. While these average error spectra are not the true error spectra, they are the best estimates available.

\subsection{AGN Identification}
\label{sec:lineratios}
In the unified model of AGNs \citep[e.g.,][]{Antonucci93, Urry95}, if our view of an AGN is unobscured, i.e. seen face-on, the spectrum contains Doppler-broadened \Ha\ and \Hb\ components from the gas clouds rapidly orbiting the supermassive black hole (SMBH) within the central parsec of the AGN. Such a system is called a ``broad line" or ``Type 1" AGN. If, instead, an AGN is observed edge-on, through the obscuring parsec-scale dusty material, only the narrow forbidden lines, from clouds hundreds to thousands of parsecs from the central engine, are visible in the spectrum. This kind of AGN is called a ``narrow line" or ``Type 2" AGN. 

\begin{figure}[htb]
\begin{center}
\includegraphics[width=0.6\textwidth,angle=90]{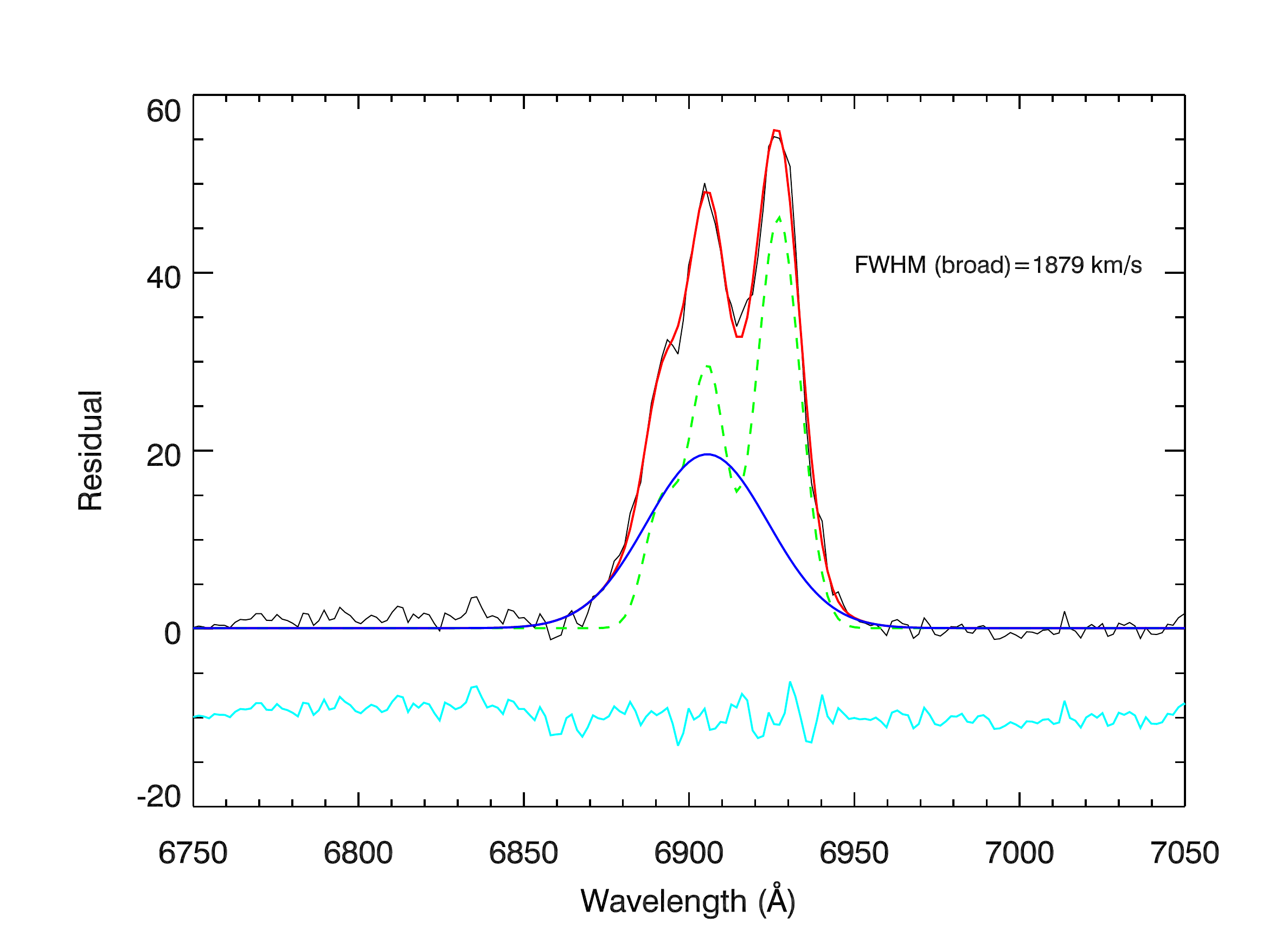}
\end{center}
\caption{Emission line profile fitting around the \NII-\Ha wavelength region for a Type 1 AGN candidate. The data is shown as a black line. The red line is a four-component Gaussian profile, while the cyan line is the residual between data and the best fit, offset by $-10$ for better visibility in the plot. The blue line shows the broad component of \Ha, and the green dashed line shows the three narrow lines, $\rm{[NII]\lambda6548}$, narrow \Ha, and $\rm{[NII]\lambda6584}$. }
\label{fig:Gaussfit}
\end{figure}

The lines we use for AGN identification are broad and narrow \Ha, and narrow \Hb, \NII~$\lambda6584$, and \OIII~$\lambda5007$\footnote{We note that the [S{\sc ii}] and [O{\sc i}] lines can be used for AGN identification in addition to, or in lieu of, \NII. However, these lines are usually weaker than \NII. Since a large fraction of our spectra have low signal-to-noise, we evaluate only the strongest lines.} After subtraction of the host galaxy contribution, the residual spectrum is fit with Gaussians to determine the full width at half maximum (FWHM) and flux of each of these four lines. For the \NII-\Ha\ complex where the lines are blended together, we use three or four Gaussians to fit all the emission components simultaneously. Figure~\ref{fig:Gaussfit} shows an example of the quadruple Gaussian fit (the two narrow \NII\ lines, broad \Ha, and narrow \Ha). The flux of the line is obtained by summing the fluxes in the wavelengths bins within $3\sigma$ of the fitted peak, where $\sigma$ is the fitted width of the Gaussian. The flux error is obtained by adding, in quadrature, the value of each bin within $3\sigma$ of the fitted peak in the error spectrum.  A broad \Ha\ line is considered to be detected if the broad \Ha\ flux divided by the flux error is more than 2.0. If the broad \Ha\ component does not meet the criterium for an emission line, the \NII-\Ha\ is refit with three Gaussians for a better determination of the narrow line fluxes. 

We identify Type 1 AGNs by the presence of a broad component of \Ha\ with S/N $\geq$ 2.0 (3.0 for SDSS) and the widely used requirement \citep[e.g.,][]{Eun17, Ho97, Schneider10, Stern12, VandenBerk06} of full width at half maximum (FWHM) $\geq$ 1000 km s$^{-1}$. Since \Hb\ is generally weaker, we do not impose any requirements for the presence of a broad \Hb\ line. However, if we find a broad \Hb\ line but not a broad \Ha\ line, we also accept that galaxy as a Type 1 AGN. There are only five such galaxies in our sample. A weak broad component of \Ha\ may be unidentified in low signal-to-noise cases.

When a broad line is not present, we use the ratios of the fluxes of the narrow forbidden lines, \OIII and \NII, to the fluxes of the narrow hydrogen Balmer lines \Hb\ and \Ha, respectively, to identify AGNs. Star formation can also excite the forbidden lines but their ratio to the Balmer lines is lower than the forbidden lines excited by AGNs. The use of these line ratios was first proposed by \citet{BPT}, and the plot of the \OIII/\Hb\ ratio versus the \NII/\Ha\ ratio is known as the BPT diagram. There are two demarcation lines usually used to separate AGNs and star forming galaxies, namely,
\begin{align*}
&\log(\rm{[OIII]}/\Hb)  > 0.61/(\log(\rm{[NII]}/\Ha) - 0.47) + 1.19 &\\
&\log(\rm{[OIII]}/\Hb)  > 0.61/(\log(\rm{[NII]}/\Ha) - 0.05) + 1.3. &
\end{align*}

The first, developed by \citet{Kewley01}, is based on theoretical modeling of maximal line ratios possible in star formation, and the second, developed by \citet{Kauffmann03}, is based on empirical studies of SDSS galaxies. Those galaxies falling above the \citet{Kewley01} line have forbidden line emission dominated by the AGN and those between the \citet{Kewley01} and \citet{Kauffmann03} lines are ``composite" galaxies with mixed AGN and star formation contributions. We use the \citet{Kauffmann03} line to identify AGNs for our catalog. If an AGN is also above the \citet{Kewley01} line, we mark it as such.

\section{Validation of Method and Cross Checks}
\label{sec:validation}
The SDSS subsample provides an ideal sample for checking our method and results. It has spectra with the highest S/N and absolute flux calibration and line measurements for the sample have been published \citep[e.g.,][]{SDSS, Greene07, Thomas13}. We also use the results from the SDSS spectra to compare with those from 6dF for the galaxies which have been observed with both instruments. This step allows us to study the effects of signal-to-noise. Since the 6dF, FAST, and CTIO spectra are of similar data quality, results from the SDSS-6dF comparison can be applied to the FAST and CTIO subsamples. 

\subsection{Validation of Type 1 AGN Identification}
\label{sec:t1valid}
We first check our broad line AGN identification. We compare our broad line AGN determination against the SDSS Seyfert 1 sample published by \citet{Greene07}. We cross match 2MRS to this sample and find 176 Type 1 AGNs that \citet{Greene07} determine to have \Ha\ FWHM $\geq$ 1000 km/s. All of them are identified as Type 1 AGNs in our analysis. Therefore, we are confident that our method correctly identifies broad line AGNs.

\subsection{Validation of Type 2 AGN Identification}
\label{sec:t2valid}
\begin{figure}[htb]
\begin{center}
\includegraphics[scale=0.7]{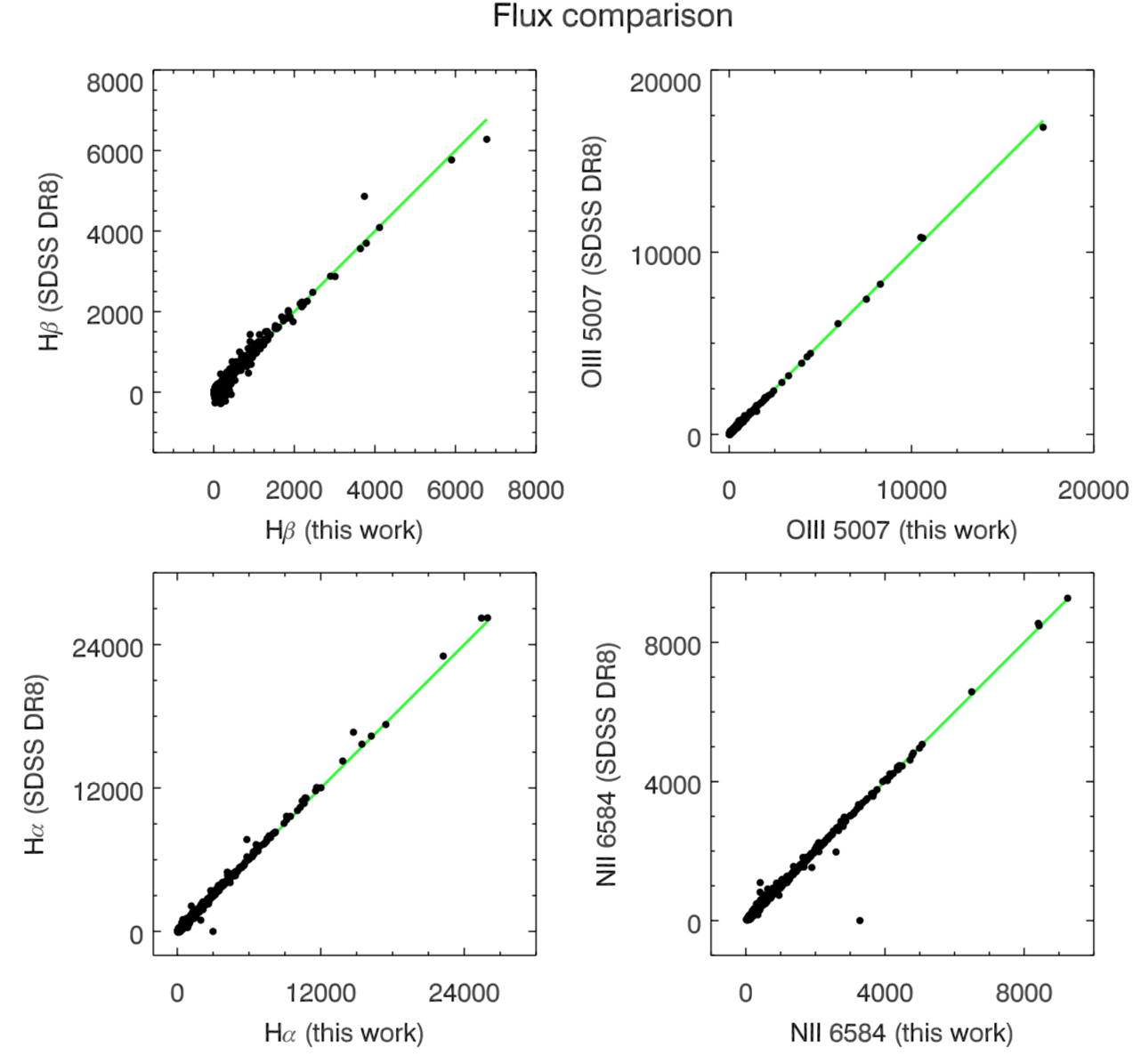}
 \end{center}
\caption{Fluxes measured in this work vs. fluxes from SDSS Data Release 8 \citep[DR8]{SDSS}. The units are 10$^{-17}$ erg cm$^{-2}$ s$^{-1}$. The fluxes correlate tightly. The green lines show a one-to-one correlation to guide the eye.}
 \label{fig:SDSSflux}
\end{figure}

In order to validate our narrow line AGN identification, we begin by checking our flux measurements. Figure~\ref{fig:SDSSflux} shows the comparison of the narrow line fluxes from SDSS Data Release 8 \citep[DR8]{SDSS} against ours, for the four lines used in AGN identification. Our measurements correlate tightly with those from SDSS. All four slopes are close to 1.0, though slightly different for each of the four lines, and the relation shows some scatter. We note that we use the MILES stellar population model (SPM) while SDSS uses the \citet{BC03} (BC03) SPM, so some differences are expected. As explained below, while BC03 was the best available model at the time of DR8, MILES templates are based on a better stellar library and give more correct results.

\begin{figure}[htb]
\begin{center}
\includegraphics[scale=0.7]{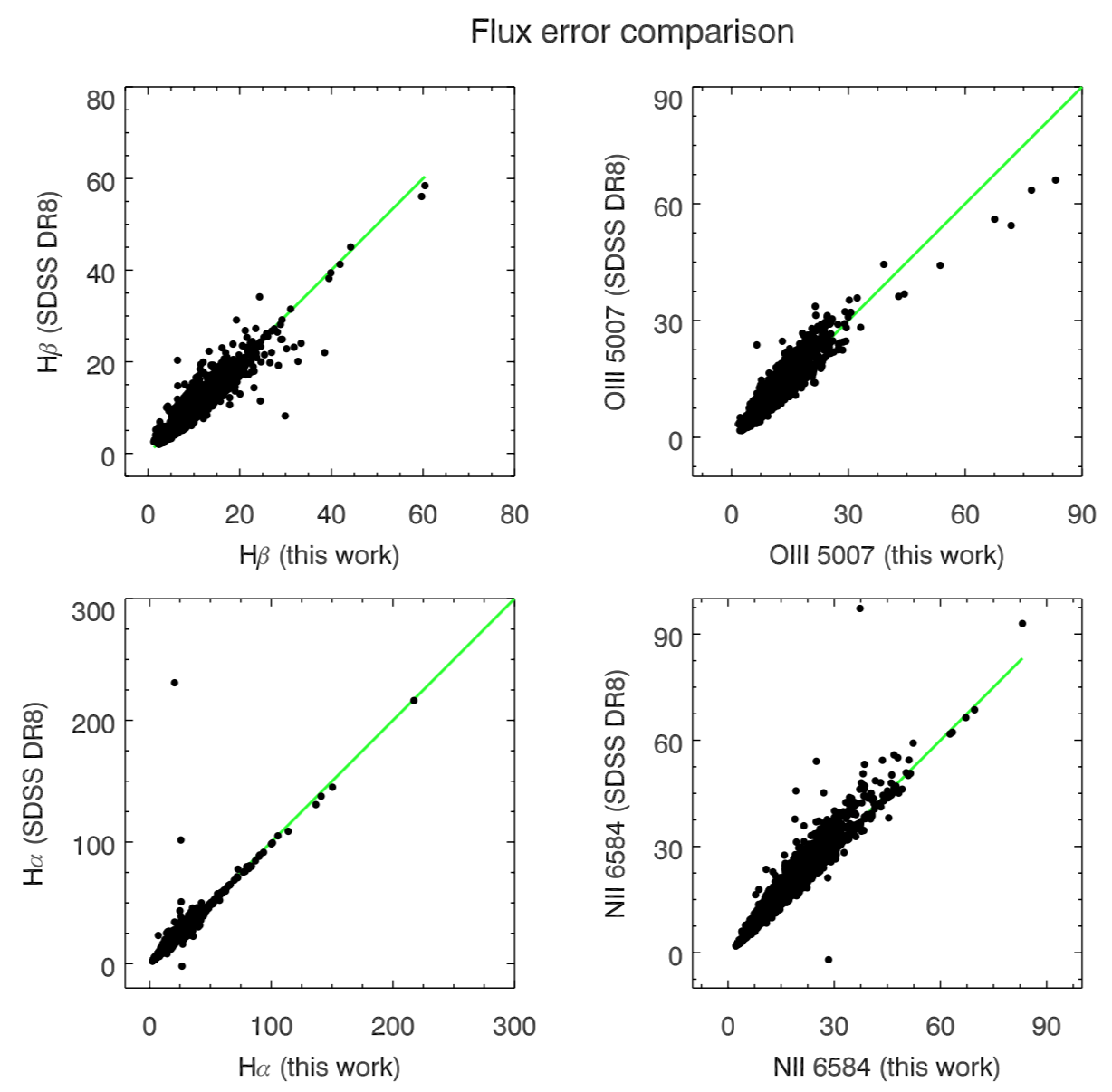}
\end{center}
\caption{Flux errors measured in this work vs. flux errors from SDSS Data Release 8 \citep[DR8]{SDSS}. The units are 10$^{-17}$ erg cm$^{-2}$ s$^{-1}$. The green lines show a one-to-one correlation to guide the eye.}
\label{fig:SDSSflux_err}
\end{figure}

Figure~\ref{fig:SDSSflux_err} shows the comparisons of the flux errors from DR8 and those we have determined. The errors show general agreement but the scatter is larger than the scatter in the fluxes. We note that these errors do not include the contributions from the galaxy contribution and emission line fitting procedures. We have performed Monte Carlo studies where we vary the noise and repeat the fitting procedure 100 times per spectrum for $\sim$20 galaxies. These studies show that the contribution from the fitting procedure could increase the error by up to a factor of two. Performing this procedure on all spectra is too computationally intensive to be practical, however the fact that both the fluxes and flux errors are consistent with those published by SDSS gives us confidence that our method is sound and that our signal-to-noise cuts are comparable to other analyses. 

\subsection{Dependence of Line Ratios on Stellar Population Models}
Our calculated line fluxes correlate well with the values published in SDSS DR8, but a more crucial check for AGN identification is whether the line ratios also agree. Our line ratios are systematically different from those of SDSS DR8 and of \citet{Thomas13}, both publicly available for the full SDSS sample. These differences are due to the different stellar population models used to subtract the host galaxy contribution in the three analyses. As detailed in Section~\ref{sec:ssps}, we use the MILES \citep{MILES} stellar population model (SPM) while SDSS DR8 uses the \citet[BC03]{BC03} SPM and \citet{Thomas13} uses only the solar metallicity single stellar population templates (SSPs) from the MILES-based \citet{MS11} SPM (MS11$_{\rm solar}$). The complete analysis has been published in our paper \citet[CZF18]{CZF}. The results, summarized below, give us confidence that the MILES templates yield line fluxes, and ratios, that are more correct than those based on BC03 and MS11$_{\rm solar}$. Equally importantly, the CZF18 analysis shows that large systematic differences can result due to a choice of stellar population models and these effects must be taken into account when constructing or using any optical AGN catalog.

\begin{figure} [htb]
\begin{center}
\includegraphics[scale=0.65]{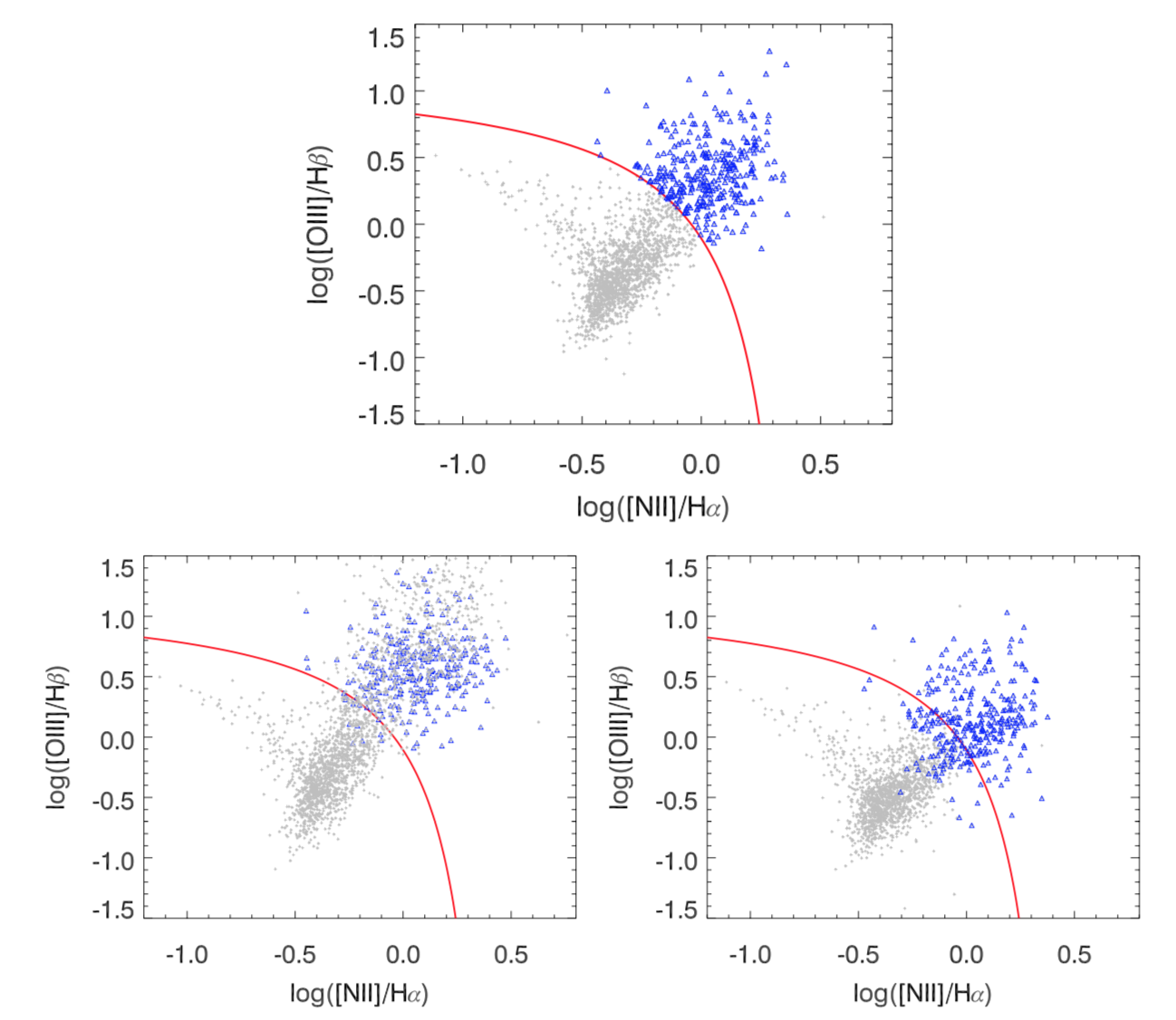}
\end{center}
\caption{Top: BPT diagram for the SDSS subsample using MILES templates to subtract the galaxy contribution from the spectrum. Star-forming galaxies are shown as grey crosses and AGNs as blue triangles. The included galaxies are required to have $\rm S/N \ge$3 for all four narrow lines used in Type 2 AGN identification. Bottom, left: BPT diagram with line ratios for the same galaxies, but using BC03 for for subtraction of host galaxy contribution. Bottom, right: BPT diagram for the same galaxies, but using MS11$_{solar}$ for subtraction of host galaxy contribution. As further explained in the text, the BC03 templates have shallower \Ha\ and \Hb\ lines due to the smaller parameter space coverage of the underlying stellar library, resulting in systematically higher line ratios. We also find that these galaxies have higher-than-solar metallicity in the MILES fits. When fitting with MS11$_{solar}$, using only the solar metallicity templates results in higher \Ha\ and \Hb\ emission, and, consequently, systematically lower line ratios. This figure is reproduced from \citet{CZF} by permission of the AAS.}
\label{fig:bpt_ssp}
\end{figure}

Figure~\ref{fig:bpt_ssp} shows the BPT diagrams for the full SDSS subsample using line ratios from this work (top), SDSS DR8 (bottom, left), and \citet{Thomas13} (bottom, right), with the \citet{Kewley01} demarcation line shown in red. It is evident that the DR8 line ratios are systematically higher than ours, shifting more galaxies into the AGN region, while the \citet{Thomas13} line ratios are systematically lower, shifting galaxies into the star forming region. The fraction of galaxies which shift from AGN to star forming or vice-versa is only a few percent for high luminosity AGNs but can be up to 50\% for AGNs with \OIII\ luminosity less than $10^{38}$ erg s$^{-1}$. In the full sample, $\sim$25\% of galaxies identified as Type 2 AGNs from the SDSS DR8 line ratios are not AGNs based on our results and $\sim$22\% of the galaxies identified as AGNs by us fall below the line when using \citet{Thomas13} line ratios. These discrepancies are due to the differences in the stellar population models.

Compared to MILES, BC03 templates are based on stellar libraries with fewer stars covering a smaller region in parameter space. In comparing the BC03 SSPs against MILES and purely theoretical SSPs \citep{g05}, BC03 templates have shallower \Ha\ and \Hb\ absorption lines, as pointed out by \citet{g05}. Thus, the \Ha\ and \Hb\ emission are systematically too small after galaxy subtraction using BC03. Consequently, the \NII/\Ha\ and \OIII/\Hb\ ratios are over-estimated, giving systematically higher AGN detection rates. 

The discrepancies with \citet{Thomas13} is more surprising at first because the stellar population templates used are also based on the MILES stellar library. However, \citet{Thomas13} only used the templates with solar metallicity to lower computing time, claiming that there is a degeneracy in age and metallicity. While it is true that there is a degeneracy in age and metallicity when fitting only colors, the degeneracy is lifted when both the continuum and absorption lines are fitted simultaneously \citep[e.g.,][]{Reichardt01}. A large majority of our best fits are with templates which have higher-than-solar metallicity. The absorption lines of solar metallicity templates are deeper than those with super-solar metallicities. Therefore, we deduce that the \Ha\ and \Hb\ emission line fluxes from \citet{Thomas13} are systematically too large and the BPT line ratios are systematically lower.

\subsection{Dependence of AGN Identification on Emission Line Signal-to-Noise Requirements}
\label{sec:linesnr}

\begin{figure}[htb]
\begin{center}
\includegraphics[scale=0.7]{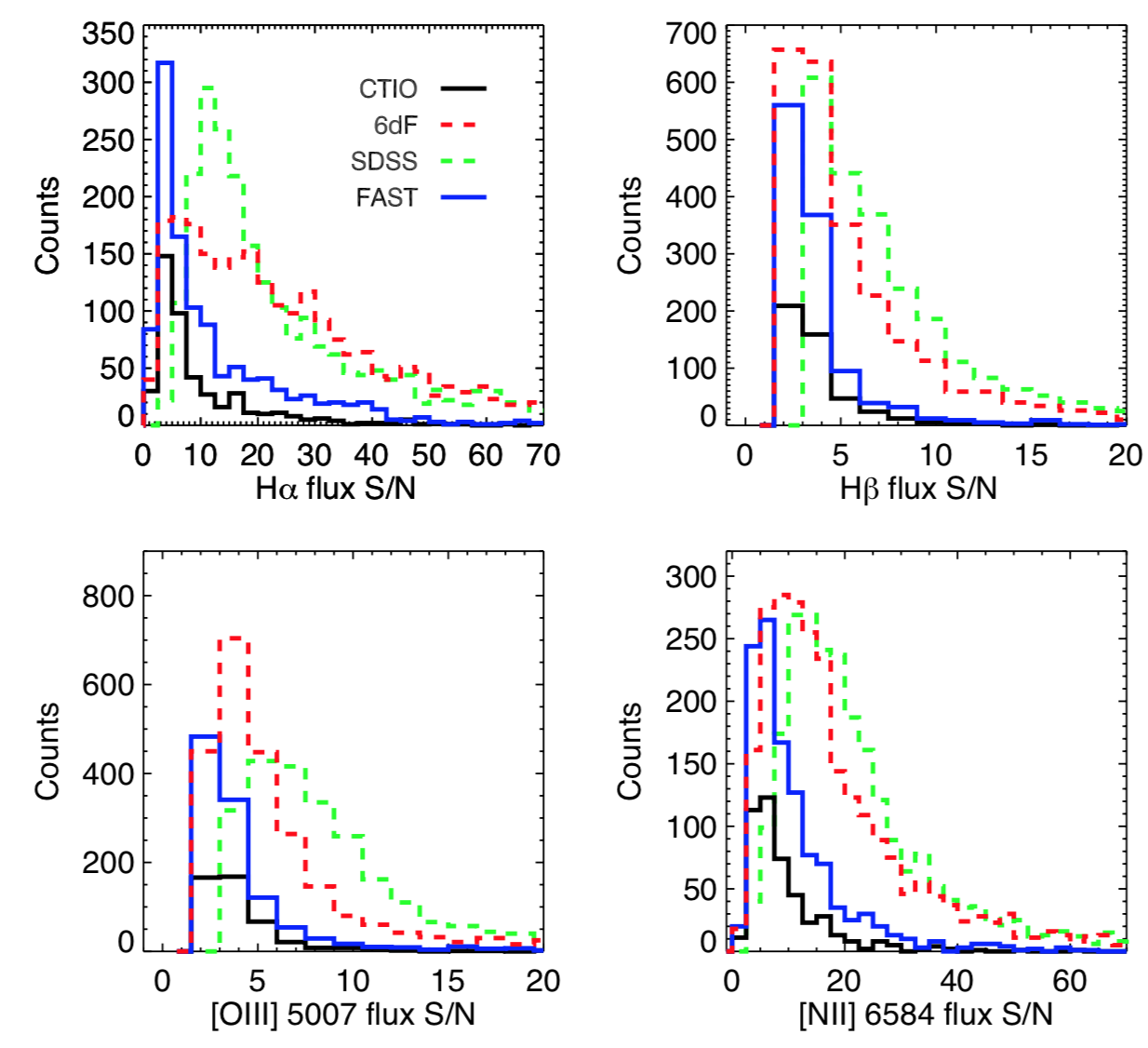}
\end{center}
\caption{S/N for line fluxes. The different subsamples have very different flux S/N values. While we require that SDSS spectra have S/N $\geq$ 3.0 for all four lines, we lower it to S/N $\geq$ 2.0 for all four lines for 6dF, FAST, and CTIO since those subsamples have lower signal-to-noise, as discussed in the text.}
\label{fig:lineSNR}
\end{figure}

As noted before, the SDSS subsample has the highest quality spectra. Having validated our method on the SDSS sample, we need to understand how the results will change when applied to noisier spectra. As seen earlier in Figure~\ref{fig:sampleSNR}, the 6dF, FAST, and CTIO spectra have lower continuum signal-to-noise (S/N) throughout the entire wavelength range, compared to SDSS spectra. The extra noise also affects the ability to detect the emission lines with confidence, and we see, in Figure~\ref{fig:lineSNR}, that 6dF, FAST, and CTIO subsamples have lower S/N for the line emission, compared to SDSS. In order for the catalog to be as homogeneous as possible, we want the AGN detection fraction for the four subsamples to be as close as possible. Since SDSS has higher S/N on average, requiring the same S/N for SDSS and the other three subsamples would necessarily result in a lower AGN detection rate for the galaxies with 6dF, FAST, and CTIO spectra. Fortunately, the SDSS footprint overlaps that of 6dF in the three stripes in the southern hemisphere. We use the 382 galaxies in 2MRS that have spectra from both SDSS and 6dF to assess the impact of line S/N on AGN detection rate. Of these, 334 galaxies have 6dF spectra outside the telluric contamination redshift range. 

\begin{figure}[htb]
\begin{center}
\includegraphics[width=0.475\textwidth]{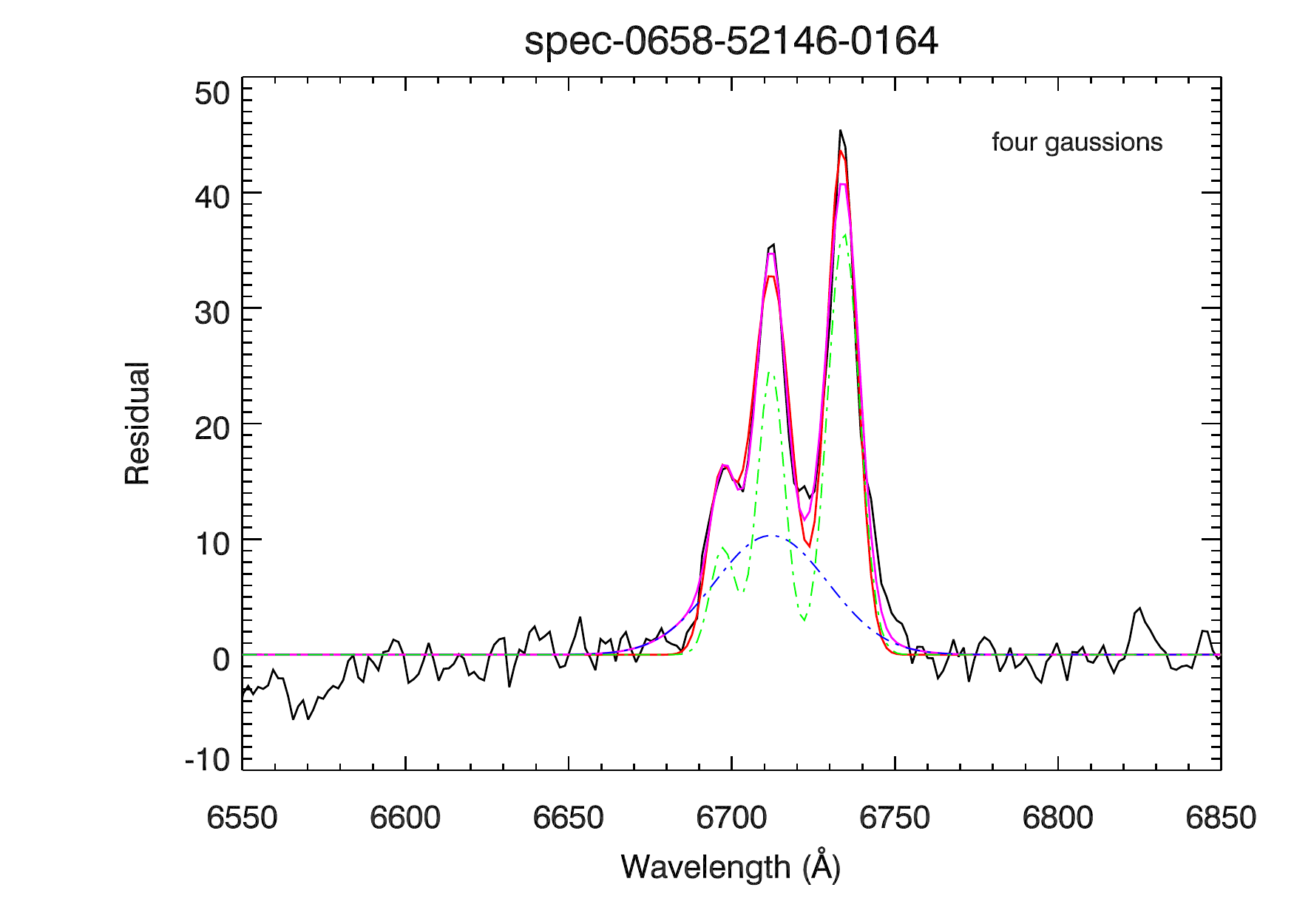}
\includegraphics[width=0.475\textwidth]{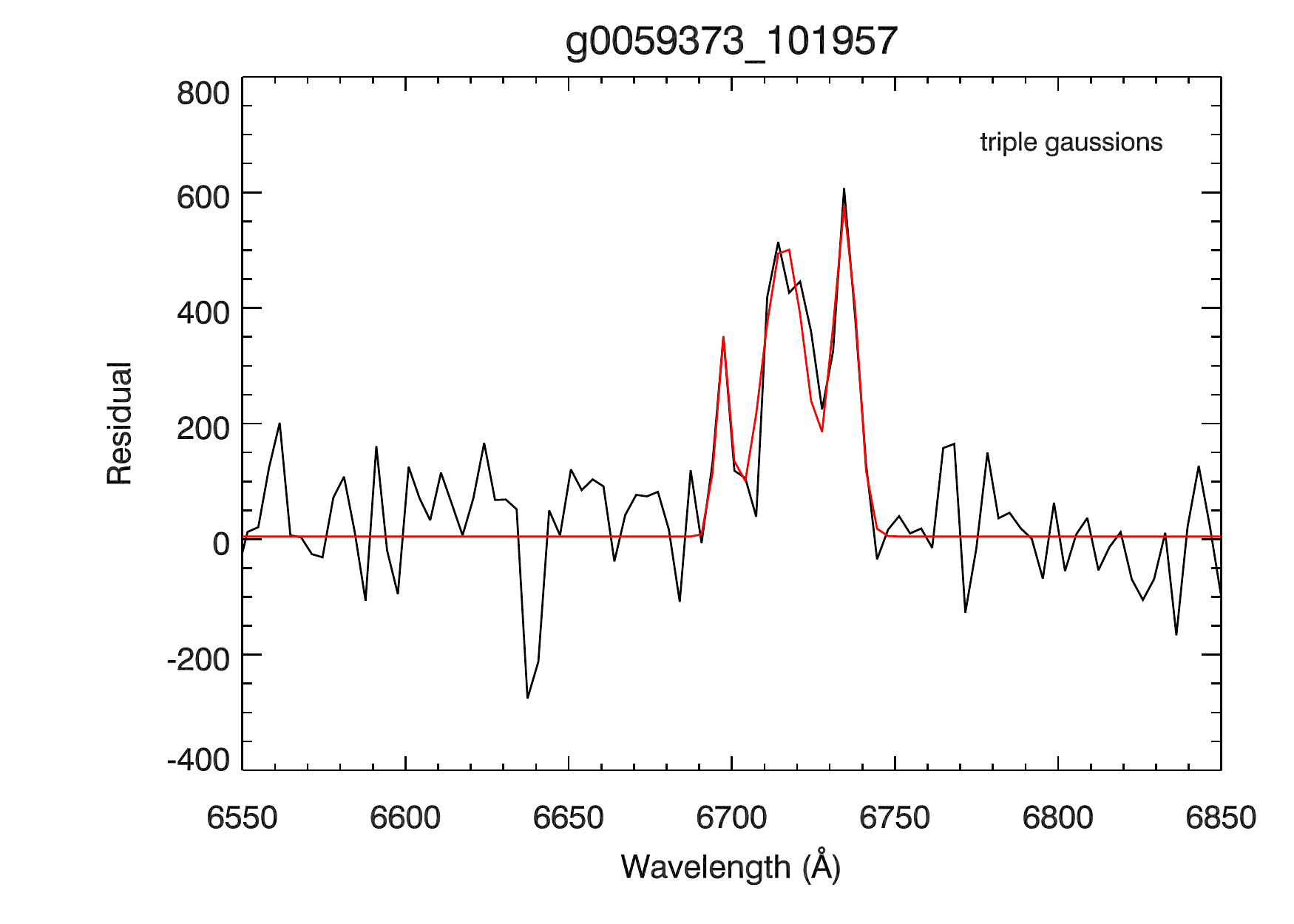}
\end{center}
\caption{SDSS spectrum (left) and 6dF spectrum (right) for the same galaxy. The broad \Ha\ is clear in the SDSS spectrum but is not detected in 6dF due to lower signal-to-noise.}
\label{fig:6dFSDSSSy1}
\end{figure}

Signal-to-noise affects the detection of both Type 1 and Type 2 AGNs. Figure~\ref{fig:6dFSDSSSy1} shows an example of a galaxy for which a broad \Ha\ component was detected in the SDSS spectrum but not the 6dF spectrum. Broad line AGNs generally have strong narrow line emission as well, and when the broad line is not evident, can often be identified as AGNs from their narrow lines. We use the sample of galaxies with SDSS and 6dF spectra to study the differences in Type 2 AGN detection rates due to line S/N.

\begin{figure}[htb]
\begin{center}
\includegraphics[scale=0.7]{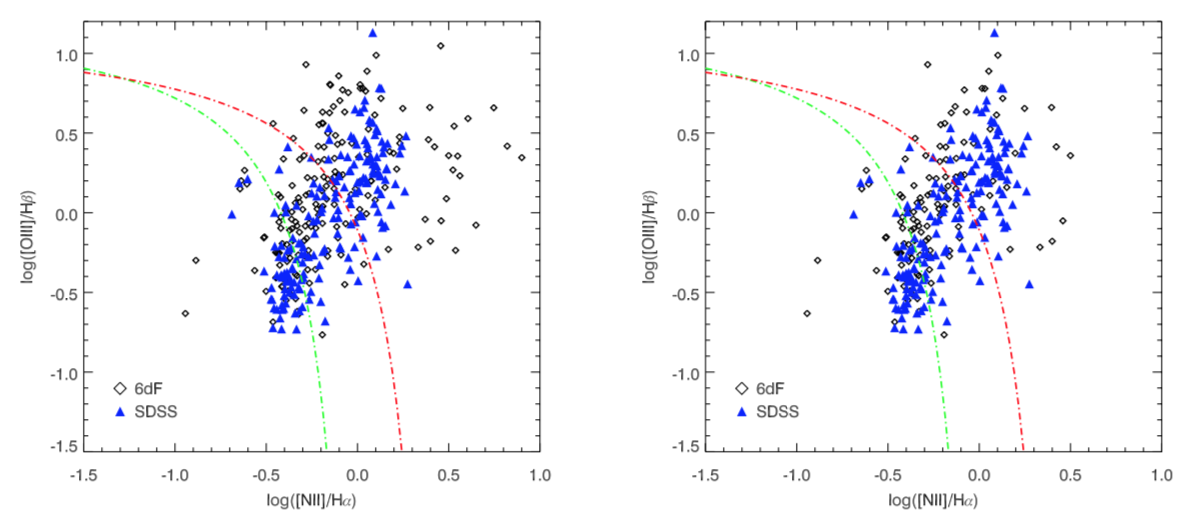}
\end{center}
\caption{Left: The BPT diagram of the galaxies with both SDSS and 6dF spectra, with all four narrow AGN diagnostic lines having S/N $\geq$ 3.0 (SDSS) and S/N $\geq$ 1.2 (6dF). The S/N cuts are set to give the same AGN fractions in both samples. The log(\NII/\Ha) ratio is systematically higher for the 6dF measurements. Right: Same plot but with the 6dF spectra required to have S/N $\geq$ 2.0 for all four lines. The systematic differences between SDSS and 6dF are greatly reduced.}
\label{fig:BPT_6dFSDSS}
\end{figure}

We determine how low the S/N cut would have to be such that the same number of galaxies would be identified as Type 2 AGNs when using 6dF spectra as when using SDSS spectra. We find that lowering the 6dF S/N cut to 1.20 for all four lines, gives the same fraction of detected Type 2 AGNs which meet the \citet{Kewley01} criteria. However, when we plot these galaxies, with spectra from both SDSS and 6dF, on the BPT diagram, we find that the \NII-\Ha\ ratios fit from the 6dF spectra are systematically higher than those from the SDSS spectra, as shown in Figure~\ref{fig:BPT_6dFSDSS}, left panel. This indicates that such a low cut on 6dF S/N does not yield reliable results. 

\begin{figure}[htb]
\begin{center}
\includegraphics[width=0.5\textwidth, angle=90]{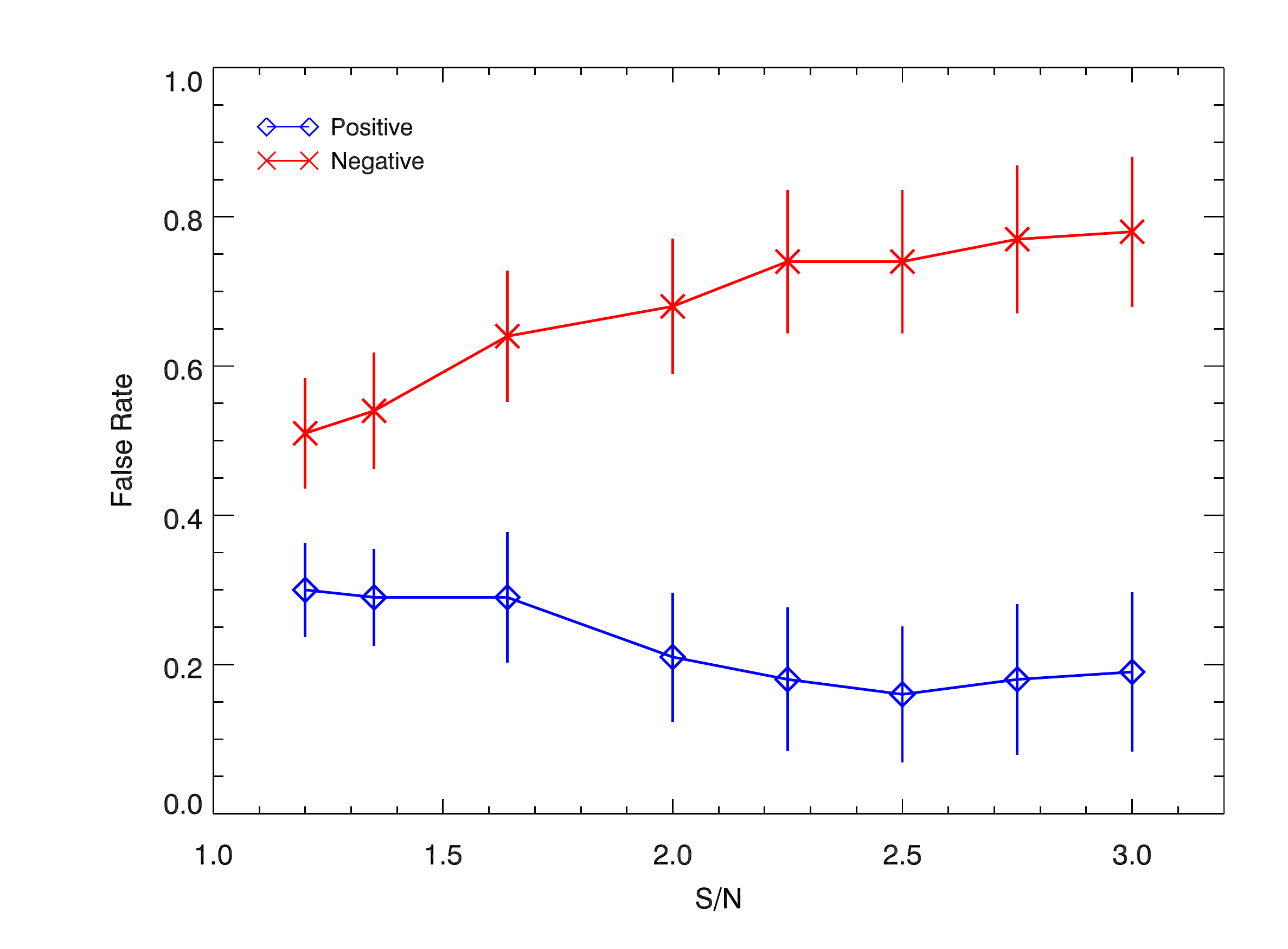}
\end{center}
\caption{The rates of false positives (blue) and false negatives (red) in AGNs identified from 6dF spectra compared to SDSS spectra, as a function of the S/N requirement for the four Type 2 AGN identification lines.} 
\label{fig:SNRcut_study}
\end{figure}

Choosing an optimal S/N cut necessarily involves a trade off in purity, i.e. minimizing false positives, and completeness, i.e. minimizing false negatives. To find the correct balance we start with a 6dF S/N of 1.2 and raise it in small increments. At each step, we calculate the rate of false positives, defined as the number of AGNs (Type 1+Type 2) identified using 6dF spectra but not SDSS spectra divided by the number of AGNs identified from 6dF spectra, and the rate of false negatives, defined as the number of AGNs (Type 1+Type 2) identified using SDSS spectra but not 6dF spectra divided by the number of AGNs identified from SDSS spectra. The results are shown in Figure~\ref{fig:SNRcut_study}. We choose a requirement of S/N $\geq$ 2.0 for all four narrow lines in the 6dF spectra as a reasonable point for the trade off between false positives and false negatives. Figure~\ref{fig:BPT_6dFSDSS}, right panel, shows the BPT diagram of AGNs with S/N $\geq$ 2.0 for all four lines in 6dF spectra, compared to the AGNs with S/N $\geq$ 3.0 using SDSS spectra.  The systematic errors in 6dF \NII/\Ha\ line ratio are significantly reduced compared to the when the 6dF spectra were only required to have S/N $\geq$ 1.2. 

\begin{figure}[htb]
\begin{center}
\includegraphics[width=0.75\textwidth]{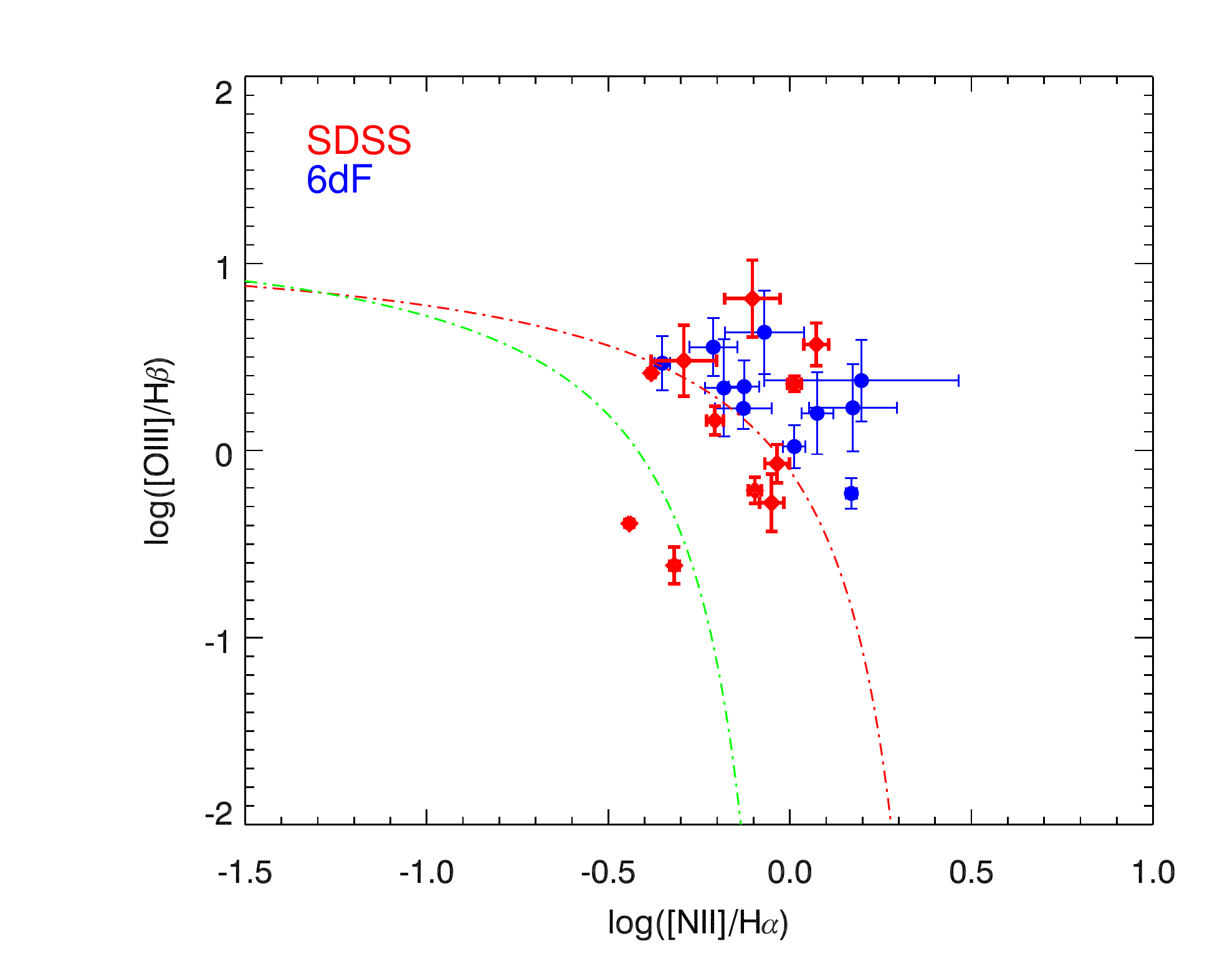}
\end{center}
\caption{BPT diagram of the AGNs identified from 6dF spectra (with S/N $\geq$ 2.0 for all four lines) but not SDSS spectra (with S/N $\geq$ 3.0 for all four lines) for the galaxies which have both 6dF and SDSS spectra. Only two of the eleven are clearly not AGNs and another is not an emission line galaxy according to its SDSS spectrum. The ones that have line ratios above the \citet{Kewley01} line using their SDSS spectra do not pass the SDSS S/N cut. In total there are 56 AGNs identified from 6dF spectra. This indicates that the false rate for AGNs identified from 6dF spectra is only $\sim$5\%.} 
\label{fig:BPT_falsepos}
\end{figure}

Using the S/N$\geq$2.0 cut, we further examine the AGNs identified from 6dF spectra but not SDSS spectra. We plot the line ratios from 6dF and SDSS spectra for these galaxies in Figure~\ref{fig:BPT_falsepos} and see that most of these are in fact above the \citet{Kewley01} line on the diagram, but they did not have S/N $\geq$ 3.0 for all four lines in the SDSS spectra. Only three out of twelve are significantly below the \citet{Kauffmann03} line. This indicates that only a few percent of the AGNs identified from the 6dF spectra with the lower S/N cut would not be identified as such from a higher quality spectrum. Since the FAST and CTIO samples have similar S/N to 6dF, we adopt the requirement of S/N $\geq$ 2.0 for all four narrow lines for those subsamples as well. We also lower the S/N cut for broad \Ha\ for 6dF, FAST, and CTIO to 2.0.

\section{The Catalog}
\label{sec:catalog}

We have constructed an AGN catalog from the 2MRS parent sample using optical lines. Our AGN catalog consists of 1929 broad line AGNs, 3607 narrow line AGNs which satisfy the \citet{Kewley01} criteria, and 6562 narrow line AGNs which satisfy the \citet{Kauffmann03} criteria. The break down of the AGNs in each spectral subsample and the corresponding AGN fractions are given in Table~\ref{tab:AGN}. The denominators for the fractions are the number of galaxies in a given subsample that have been processed (i.e. telluric free and outside the Galactic Plane). The inhomogeneities in these AGN fractions are due to the differences in spectral quality between the subsamples as further discussed in Section~\ref{sec:detectionrates}. 

\begin{table}
 \caption{AGN Numbers and Fractions}
 \label{tab:AGN}
 \begin{center}
\begin{tabular}{|l|r|r|r|}
\hline
  \multicolumn{1}{|c|}{} &
  \multicolumn{1}{c|}{Type 1} &
  \multicolumn{1}{c|}{Type 2 K01} &
  \multicolumn{1}{c|}{Type 2 K03} \\
\hline
  6dF & 877 (8.47$\pm$0.30\%) & 1088 (10.51$\pm$0.33\%) & 2495 (24.09$\pm$0.54\%) \\
  SDSS & 811 (11.47$\pm$0.43\%) & 1511 (21.38$\pm$0.61\%) & 2455 (34.73$\pm$0.81\%) \\
  FAST & 137 (2.18$\pm$0.19\%) & 714 (11.39$\pm$0.45\%) & 1145 (18.26$\pm$0.59\%) \\
  CTIO & 104 (3.65$\pm$0.36\%) & 294 (10.31$\pm$0.63\%) & 467 (16.38$\pm$0.82\%) \\
  Total & 1929 (7.27$\pm$0.17\%) & 3607 (13.59$\pm$0.24\%) & 6562 (24.72$\pm$0.34\%) \\
\hline\end{tabular}
\end{center}
  \begin{tablenotes}
      \small
      \item The table lists the number of AGNs in each subsample as well as the AGN fraction and error on the fraction, relative to the number of spectra in our final sample, for Type 1 (broad line), Type 2 (narrow line) according to the \citet{Kewley01} (K01) criteria, and Type 2 according to the \citet{Kauffmann03} (K03) criteria. The AGN fractions are not the same for the different subsamples due to the differences in signal-to-noise, spectral resolution, and spectral sampling of each subsample. As expected, the SDSS sample has the highest AGN fractions. For Type 1 AGN, 6dF has the second highest, and FAST and CTIO have similar fractions while all three have similar fractions for Type 2 AGN. The discrepancies in Type 1 AGN fractions between 6dF, FAST, and CTIO are due to spectral resolution, as discussed in Section~\ref{sec:sy1sy2ratio} and shown in Figure~\ref{fig:sy1sy2ratio}.
    \end{tablenotes}
\end{table}

\begin{figure}[htb]
\begin{center}
\includegraphics[scale=0.7]{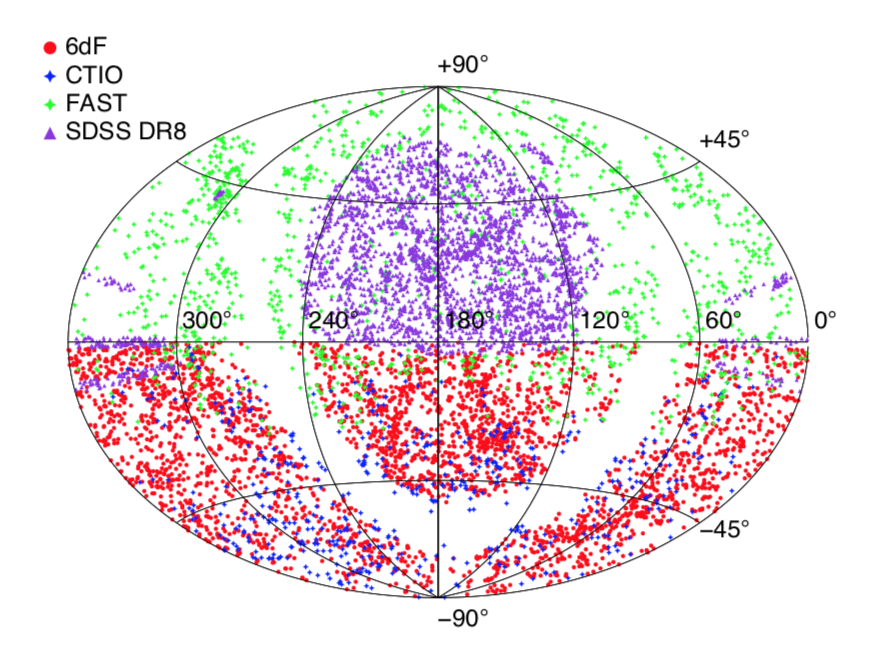}
\end{center}
\caption{The sky distribution of the AGN catalog (broad line AGNs and the narrow line AGNs satisfying the \citet{Kauffmann03} criteria). }
\label{fig:AGNskyplot}
\end{figure}

\begin{figure}[htb]
\begin{center}
\includegraphics[scale=0.7]{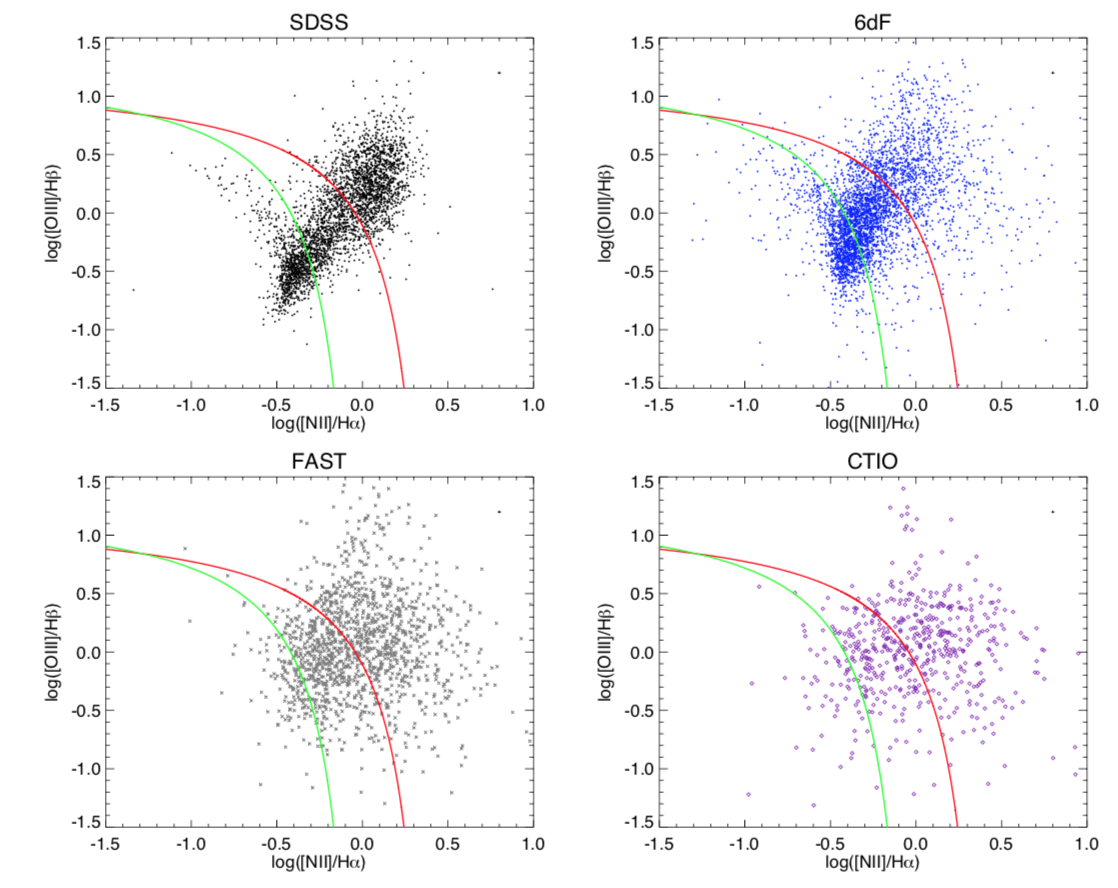}
\end{center}
\caption{The BPT diagrams for the emission line galaxies (S/N $\geq$ 3.0 for all four lines if from SDSS and S/N $\geq$ 2.0 for all four lines if from 6dF, FAST, or CTIO) for each subsample. The green and red lines denote the \citet{Kauffmann03} and \citet{Kewley01} criteria for narrow line AGNs, respectively.}
\label{fig:AGNBPT}
\end{figure}

The distribution of the AGNs in our catalog across the sky and the BPT diagrams of the emission line galaxies in each of the four subsamples are shown in Figures~\ref{fig:AGNskyplot} and~\ref{fig:AGNBPT}, respectively. As expected, we see more AGNs in the SDSS and 6dF regions than in regions covered only by FAST. The BPT diagrams shows a larger scatter for the 6dF, FAST, and CTIO galaxies due to the lower line S/N cut. 

\begin{figure}[thb]
\begin{center}
\includegraphics[scale=0.7]{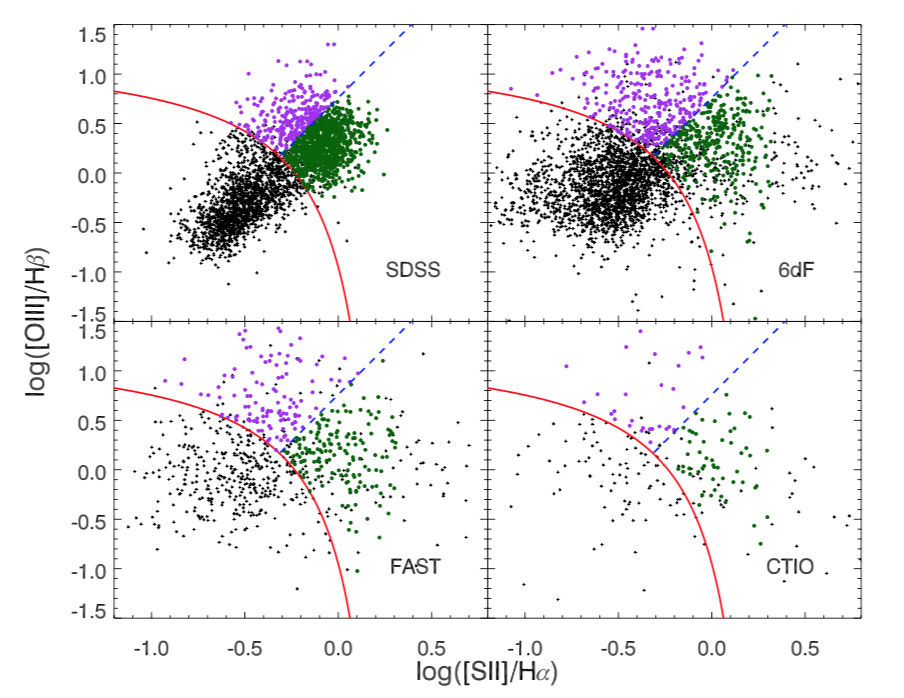}
\end{center}
\caption{The Type 2 AGNs in each subsample of our catalog, separated into Seyferts (purple) and LINERs (green) using the \citet{Kewley06} criteria. Although there are more LINERs than Seyferts in each subsample, SDSS has the highest LINER to Seyfert ratio. LINERs have weaker emission lines, so they are preferentially lost when spectral signal-to-noise decreases. Note that we use the \OIII/\Hb\ vs. \NII/\Ha\ BPT diagram to identify AGNs, so some emission line galaxies which do not qualify as AGNs in the main diagram appear in the AGN region in the \OIII/\Hb\ vs. [SII]/\Ha\ diagram.}
\label{fig:SyLINER}
\end{figure}

For the Type 2 AGNs in our catalog, we further delineate Seyferts and low-ionization nuclear emission line regions (LINERs) using the \citet{Kewley06} criteria. The Seyferts and LINERs in each subsample are shown in Figure~\ref{fig:SyLINER}. Although LINERs outnumber Seyferts in every subsample, SDSS, which has the highest S/N, has the highest LINER to Seyfert ratio. We infer that this is because LINERs have weaker emission lines and therefore, are preferentially lost in spectra with lower S/N. To test this hypothesis, we determined the LINER to Seyfert ratios with different S/N requirements and found that, as expected, the ratio decreases when the S/N is required to be higher. Note that we use the \OIII/\Hb\ vs. \NII/\Ha\ BPT diagram to identify AGNs, so some emission line galaxies which do not qualify as AGNs in the main diagram appear in the AGN region in the \OIII/\Hb\ vs. [SII]/\Ha\ diagram.

The details of our AGN identification method are described in Section~\ref{sec:AGNID}. We reiterate the selection criteria here for convenience. We require a galaxy to satisfy the following requirements to be classified as an AGN:

\begin{itemize}
\item Spectrum from SDSS, 6dF, FAST, or CTIO,
\item Spectrum does not have telluric contamination (galaxies in the range $0.0407 < z < 0.0511$ are excluded for 6dF, FAST, and CTIO subsamples)\footnote{There are 39 galaxies in the telluric contaminated redshift range which show broad \Ha\ emission wider than the telluric contamination. We include them in our catalog as Type 1 AGNs.},
\item The galaxy is outside the Galactic Plane ($|b| \geq 10^\circ$),
\item The reduced $\chi^2$ from the SSP fitting must be less than 2.55, 6.05, 3.75, and 7.15, for SDSS, 6dF, FAST, and CTIO, respectively, keeping 99\% of the subsamples,
\item Broad line AGNs (Type 1):
\begin{itemize}
\item Broad \Ha\ S/N $\geq$ 3.0 for SDSS, $\geq$ 2.0 for 6dF, FAST, and CTIO,
\item FWHM of \Ha\ $\geq$ 1000 km/s.\footnote{There are 5 galaxies which show FWHM \Hb\ $\geq$ 1000 km/s but not broad \Ha. We include them in our catalog as Type 1 AGNs.}
\end{itemize}
\item Narrow line AGNs (Type 2):
\begin{itemize}
\item All four narrow lines have S/N $\geq 3.0$ for SDSS, $\geq$ 2.0 for the other subsamples,
\item The line ratios satisfy the \citet{Kauffmann03} criteria. 
\end{itemize}
\end{itemize}
The AGNs which satisfy the more stringent \citet{Kewley01} criteria, are marked as such and form a subset of the catalog. 

A sample of the catalog is shown in Table~\ref{tab:catalog}. We provide the following properties of the AGNs:
\begin{itemize}
\item {\bf TMID}: 2MASS ID, a combination of RA and Dec in sexagesimal units.
\item {\bf RA}: Right ascension in degrees.
\item {\bf DEC}: Declination in degrees.
\item {\bf V}: Velocity in km/s, heliocentric reference frame, i.e. corrected for the motion of the Earth around the Sun.
\item {\bf CAT}: Spectral source, 6=6dF, S=SDSS, F=FAST, C=CTIO.
\item {\bf Type}: AGN Type, T1=Type 1, K01=Type 2 satisfying \citet{Kewley01} criteria, K03 = Type 2 satisfying \citet{Kauffmann03} criteria.
\item {\bf S/N (4660-4810\AA)}: Signal-to-noise ratio of the spectrum near the \Hb\ line. Since \Hb\ is the weakest of the four AGN identification lines, this is the measure of spectral quality which affects the ability to detect narrow emission lines for Type 2 AGN identification.
 \item {\bf S/N (6350-6540\AA)}: Signal-to-noise ratio of the spectrum near the \Ha\ line, the measure of spectral quality which affects the ability to detect the broad \Ha\ line for Type 1 AGN identification.
\item {\bf $M_K$}: Extinction corrected $K_s$ isophotal magnitude, this can be used to make a volume limited parent sample of galaxies.
\item {\bf SSP $\chi^2$}: The reduced $\chi^2$ from SSP fitting.
\end{itemize}

\begin{table}
 \caption{The AGN Catalog (Illustrative Extract)}
 \label{tab:catalog}
\footnotesize
\begin{center}
\begin{tabular}{|l|r|r|r|l|l|r|r|r|r|}
\hline
 \multicolumn{1}{|c|}{TMID} &
  \multicolumn{1}{c|}{RA (deg)} &
  \multicolumn{1}{c|}{DEC (deg)} &
  \multicolumn{1}{c|}{V (km/s)} &
  \multicolumn{1}{c|}{CAT} &
  \multicolumn{1}{c|}{Type} &
  \multicolumn{1}{c|}{S/N } &
  \multicolumn{1}{c|}{S/N } &
  \multicolumn{1}{c|}{$M_K$} &
  \multicolumn{1}{c|}{SSP $\chi^2$} \\
 \multicolumn{1}{|c|}{ } &
  \multicolumn{1}{c|}{ } &
  \multicolumn{1}{c|}{ } &
  \multicolumn{1}{c|}{ } &
  \multicolumn{1}{c|}{ } &
  \multicolumn{1}{c|}{ } &
  \multicolumn{1}{c|}{ (4660-4810\AA)} &
  \multicolumn{1}{c|}{ (6350-6540\AA)} &
  \multicolumn{1}{c|}{ } &
  \multicolumn{1}{c|}{ } \\

\hline
  00001131-0509313 & 0.04709 & -5.15876 & 5713 & 6 & K03 & 13.61 & 17.96 & 11.65 & 1.26\\
  00001825-4729226 & 0.07596 & -47.48958 & 8191 & 6 & K01 & 24.90 & 31.86 & 11.61 & 0.99\\
  00003564-0145472 & 0.14849 & -1.76318 & 7322 & 6 & K03 & 20.87 & 30.10 & 11.48 & 1.569\\
  00011378-4400426 & 0.3074 & -44.01183 & 11681 & 6 & K03 & 19.17 & 20.76 & 11.30 & 0.92\\
  00012334+4733537 & 0.34737 & 47.56496 & 5237 & F & K01 & 8.38 & 12.84 & 11.67 & 1.37\\
  00013605-1444548 & 0.4002 & -14.74867 & 11296 & 6 & K01 & 23.96 & 37.78 & 11.23 & 1.14\\
  00015583-2737382 & 0.48258 & -27.62723 & 8493 & 6 & K01 & 14.25 & 25.54 & 10.32 & 2.07\\
  00020386-3328023 & 0.51617 & -33.46728 & 8675 & 6 & K03 & 35.16 & 45.13 & 11.24 & 1.37\\
  00023480-0342386 & 0.64505 & -3.71072 & 6445 & 6 & T1 & 22.43 & 39.86 & 10.98 & 1.84\\
  00024429-5344549 & 0.6845 & -53.74858 & 10459 & 6 & K01 & 28.08 & 37.27 & 11.54 & 1.03\\
  00024517-3213361 & 0.68832 & -32.22675 & 8041 & 6 & K03 & 24.81 & 32.28 & 11.734 & 0.75\\
  00024862-0336216 & 0.70269 & -3.60602 & 6232 & 6 & T1 & 16.61 & 26.91 & 11.71 & 1.25\\
  00031064-5444562 & 0.79429 & -54.74892 & 9767 & C & K01 & 34.38 & 46.37 & 10.34 & 1.65\\
  00031127+1557563 & 0.79697 & 15.96568 & 11218 & S & K01 & 38.50 & 54.41 & 10.65 & 1.09\\
  00031764+7028152 & 0.82361 & 70.47093 & 7266 & F & K03 & 1.97 & 6.02 & 11.38 & 1.26\\
  06501743-3805136* & 102.57276 & -38.0871 & 9008 & C & T1 & 11.94 & 13.70 & 11.02 & 1.53\\
  07432627-6146185* & 115.85938 & -61.77181 & 10453 & C & K03 & 20.51 & 28.02 & 10.79 & 1.19\\
  09153810-8646005*\# & 138.90889 & -86.76678 & 5081 & C & K01 & 28.45 & 44.70 & 9.88 & 8.84\\
  09302571-6502042* & 142.607 & -65.03441 & 6244 & C & K01 & 10.92 & 17.37 & 11.15 & 1.72\\
  17485834-0203109* & 267.24316 & -2.05299 & 8492 & F & K01 & 9.27 & 17.77 & 11.06 & 2.58\\
\hline\end{tabular}
\end{center}
  \begin{tablenotes}
      \small
      \item TMID: Two MASS ID, a * indicates that the spectrum did not have an accompanying error spectrum and a \# indicates that ``local fitting" was used to estimate the galaxy contribution; RA: right ascension; DEC: declination; V: velocity; CAT: Spectral source, 6=6dF, S=SDSS, F=FAST, C=CTIO; Type: AGN type, T1=Type 1, K01=Type 2 with \citet{Kewley01} criteria, K03 = Type 2 with \citet{Kauffmann03} criteria; S/N (4660-4810\AA): Signal-to-noise ratio of the spectrum near the \Hb\ line; S/N (6350-6540\AA): Signal-to-noise ratio of the spectrum near the \Ha\ line; $M_K$: Extinction corrected $K_s$ isophotal magnitude; SSP $\chi^2$: Reduced $\chi^2$ from the SSP fitting. Table 3 is published in its entirety in the electronic edition of the {\it Astrophysical Journal}.  A portion is shown here for guidance regarding its form and content. The full table also includes the emission line widths, fluxes, and flux errors of all the AGNs in the catalog.
      \end{tablenotes}  
\end{table}

We also provide an extended AGN catalog which includes the FWHM and S/N of the broad \Ha\ line, used to identify Type 1 AGNs, and the fluxes, flux errors, and S/N of the four narrow lines used in identification of Type 2 AGNs. In addition, we provide a separate catalog of ``emission line galaxies", where all four narrow lines used in Type 2 AGN identification have been detected with S/N $\geq$ 1.0. For the emission line galaxy catalog, we list the TMID, RA, Dec, velocity, spectral source, $K_s$ isophotal magnitude, S/N of the spectrum near the \Hb\ line, SSP reduced $\chi^2$, and the fluxes, flux errors, and S/N of the four narrow lines. 

The line measurements we provide allow the user to construct a customized AGN catalog using different AGN selection criteria. \textbf{\textit{Note that only the SDSS spectra have absolute flux calibration. The fluxes provided for 6dF, FAST, and CTIO can only be used as ratios and not as actual flux measurements.}}

\section{AGN Detection Rates}
\label{sec:detectionrates}

In Section~\ref{sec:linesnr}, we discussed lowering the line signal-to-noise (S/N) requirement for the 6dF, FAST, and CTIO spectra relative to the SDSS spectra in order to get closer AGN detection rates. Unfortunately, we cannot lower it until the rates are equal since that introduces too many false positive AGN identifications and systematically higher \NII-\Ha\ ratios. Consequently, the AGN detection rate is lower in the 6dF, FAST, and CTIO subsamples compared to the SDSS subsample, as seen in Table~\ref{tab:AGN}. In this section, we study the effects of signal-to-noise and spectral resolution on AGN detection rates and Type 1 to Type 2 AGN ratios. 

\subsection{Impact of Spectral Resolution on the Type 1 to Type 2 AGN Ratio}
\label{sec:sy1sy2ratio}
\begin{figure}[htb]
\begin{center}
\includegraphics[width=0.75\textwidth]{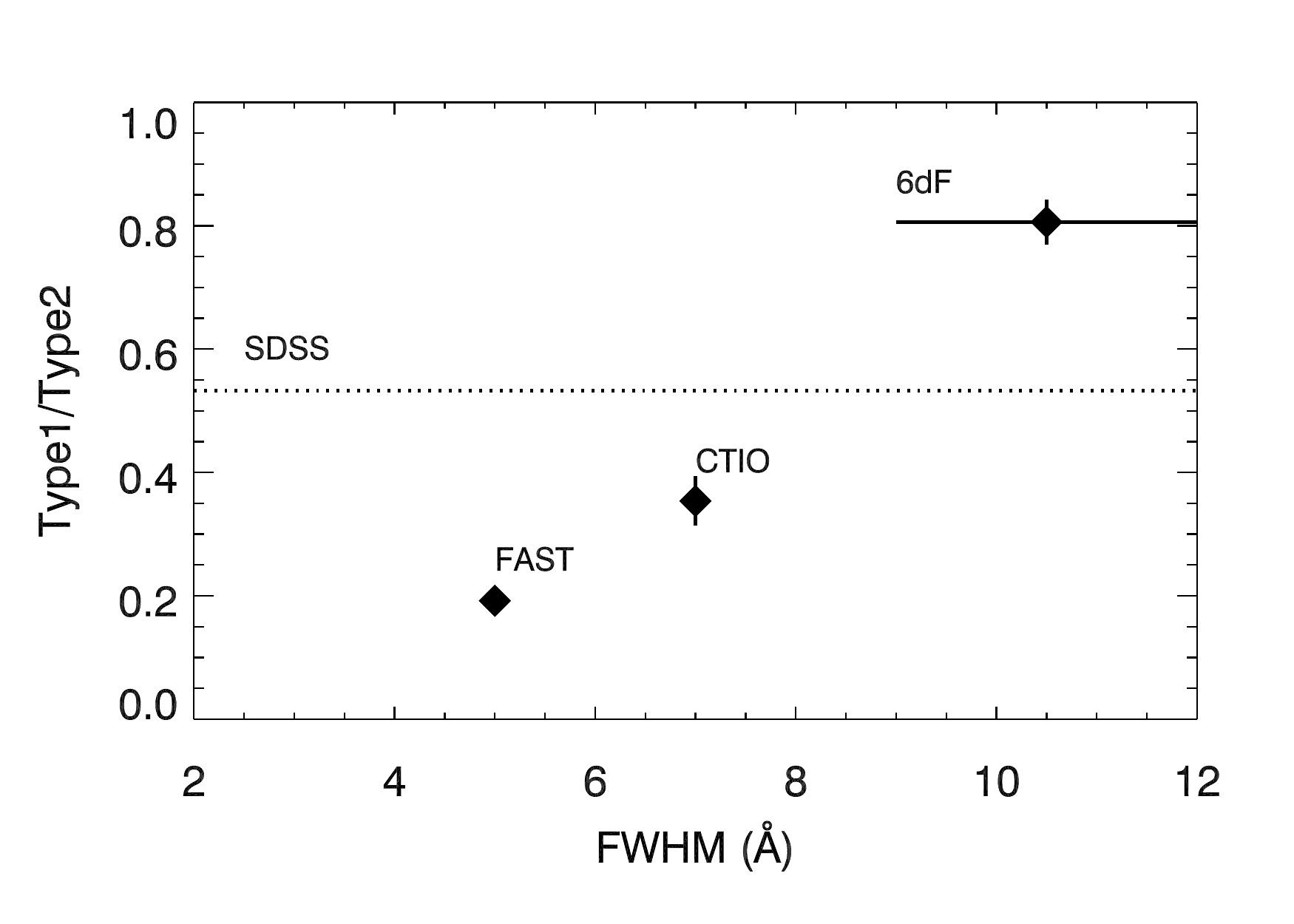}
\end{center}
\caption{The Type 1 to Type 2 AGN ratio vs spectral resolution. The dotted line shows the value for SDSS which has higher signal-to-noise spectra and a wavelength dependent resolution. We find that the Type1 to Type 2 ratio increases with lower spectral resolution, i.e. higher FWHM. This is because wider spectral bins effectively act as spectral smoothing, making it more likely to detect the broad \Ha\ which can be tens of \AA\ wide. In the wavelength range of \Ha\ and \NII, 6dF has spectral resolution $\sim$3-4 times worse than SDSS. Consequently, although the detection rate for both Type 1 and Type 2 AGNs are lower for 6dF, compared to SDSS, the Type 2 AGNs are preferentially lost at a much higher rate, leading to a higher Type 1 to Type 2 ratio in the 6dF subsample.}
\label{fig:sy1sy2ratio}
\end{figure}

For Type 2 AGNs, especially those which meet the \citet{Kewley01} criteria, the detection rates for the 6dF, FAST, and CTIO subsamples are similar, roughly half of the SDSS rate. The Type 1 AGN detection rates, however, are very different for the three lower S/N subsamples. Figure~\ref{fig:sy1sy2ratio} shows the Type 1 to Type 2 AGN ratios for the different subsamples. The SDSS value, the benchmark, is shown as a dotted line\footnote{Since SDSS spectra have higher S/N and the SDSS spectra have wavelength dependent spectral bin width to keep the wavelength uncertainty to wavelength ratio across the spectrum, we do not plot the SDSS ratio vs. spectral resolution along with the other samples.}. We find that the Type 1 to Type 2 ratio increases with spectral bin width, i.e. as resolution decreases, between FAST, CTIO, and 6dF. Since the Type 2 detection rates are roughly the same for the three subsamples, this trend is driven by Type 1 detection rates. This is due to the fact that wider spectral bins effectively act as spectral smoothing, increasing the signal-to-noise of the broad (but low) \Ha\ emission, which can be tens of \AA\ wide. In the wavelength range of \Ha\ and \NII, 6dF has spectral resolution $\sim$3-4 times worse than SDSS. Consequently, although the detection rate for both Type 1 and Type 2 AGNs are lower for 6dF, compared to SDSS, the Type 2 AGNs are preferentially lost at a much higher rate. This leads to a higher Type1 to Type 2 AGN ratio in the 6dF subsample.

\newpage
\subsection{Impact of Spectral Signal-to-Noise on AGN Detection Rates}
\begin{figure}[p]
\begin{center}
\includegraphics[width=0.45\textwidth,angle=0]{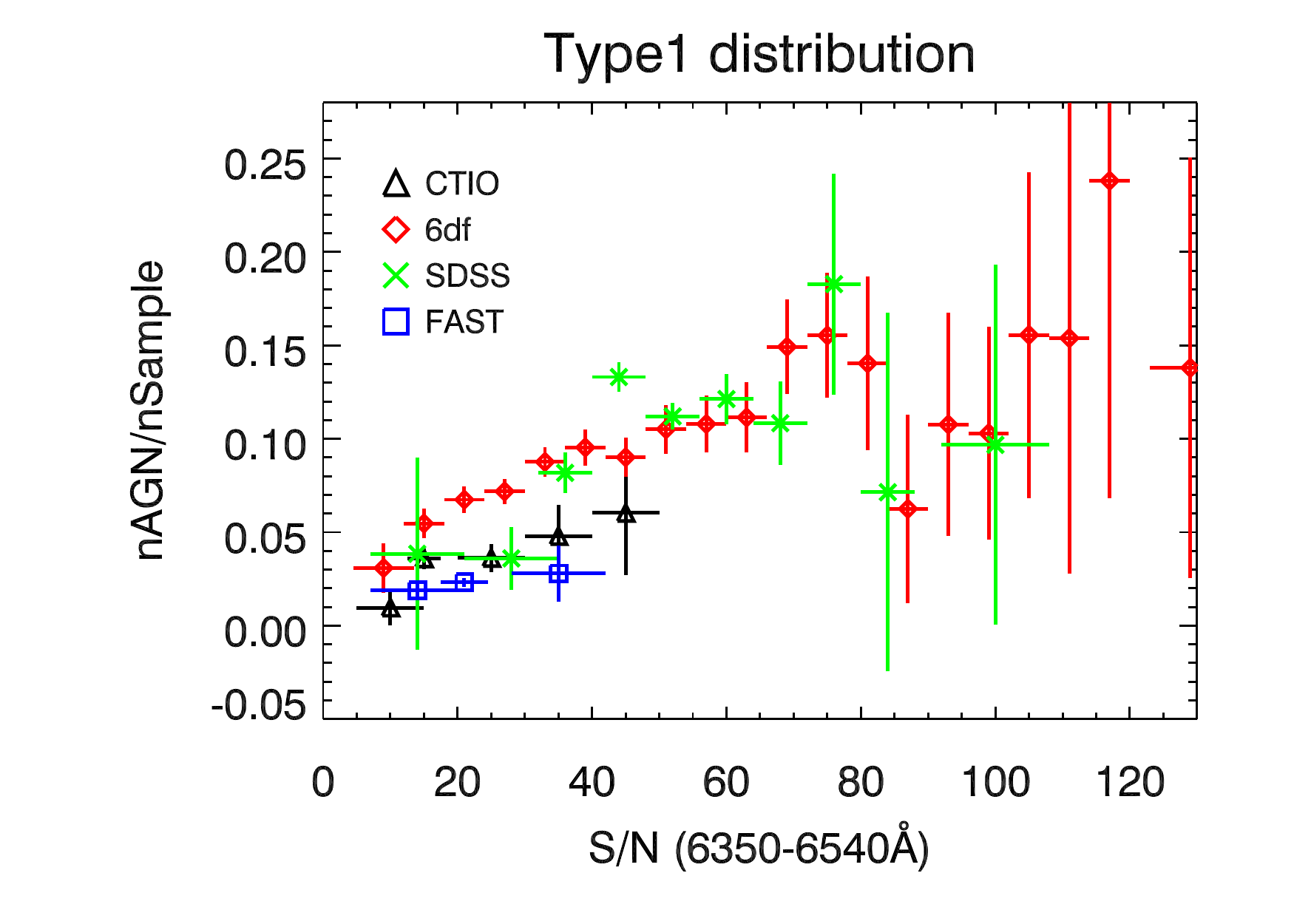}
\includegraphics[width=0.45\textwidth,angle=0]{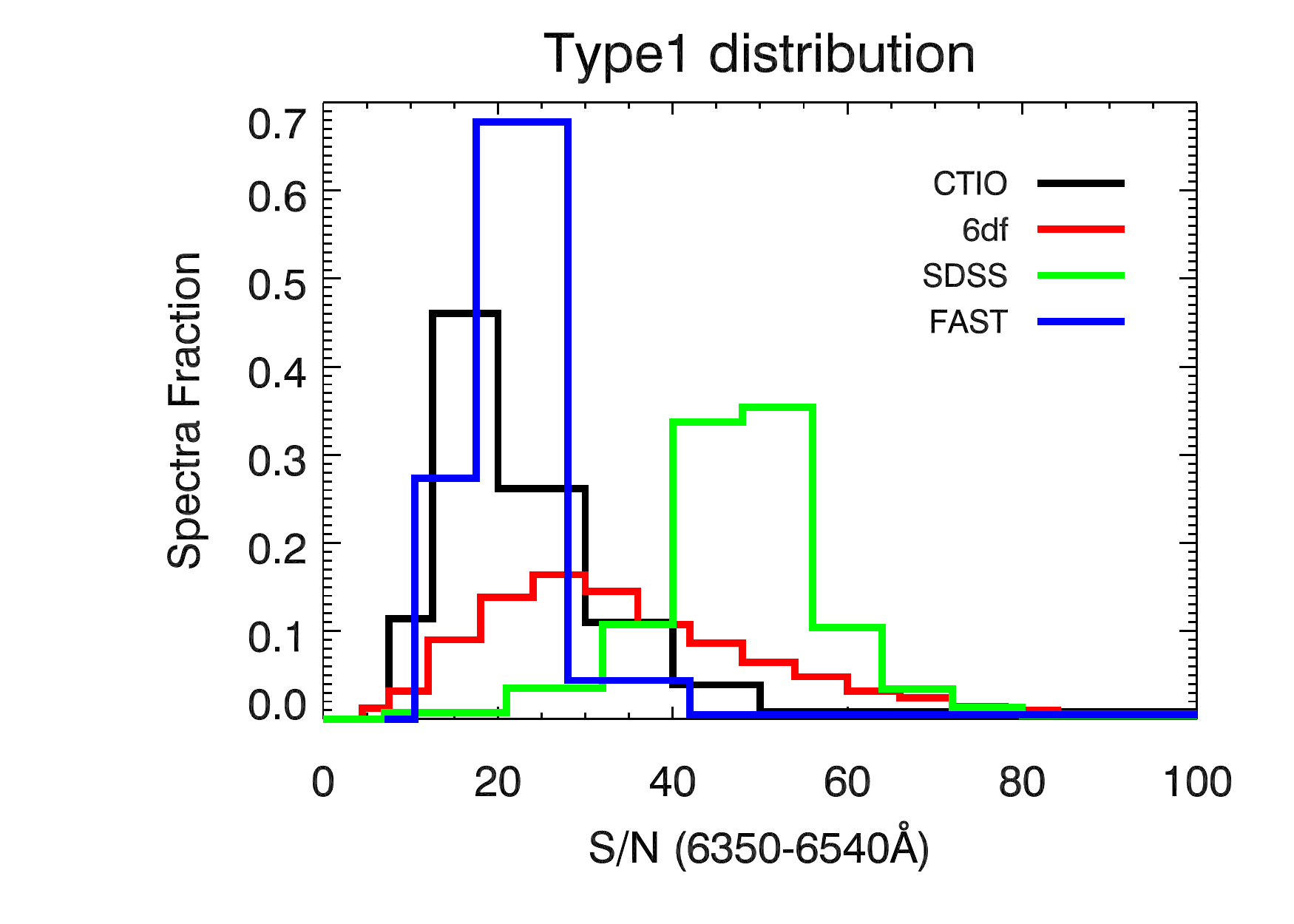}
\includegraphics[width=0.45\textwidth,angle=0]{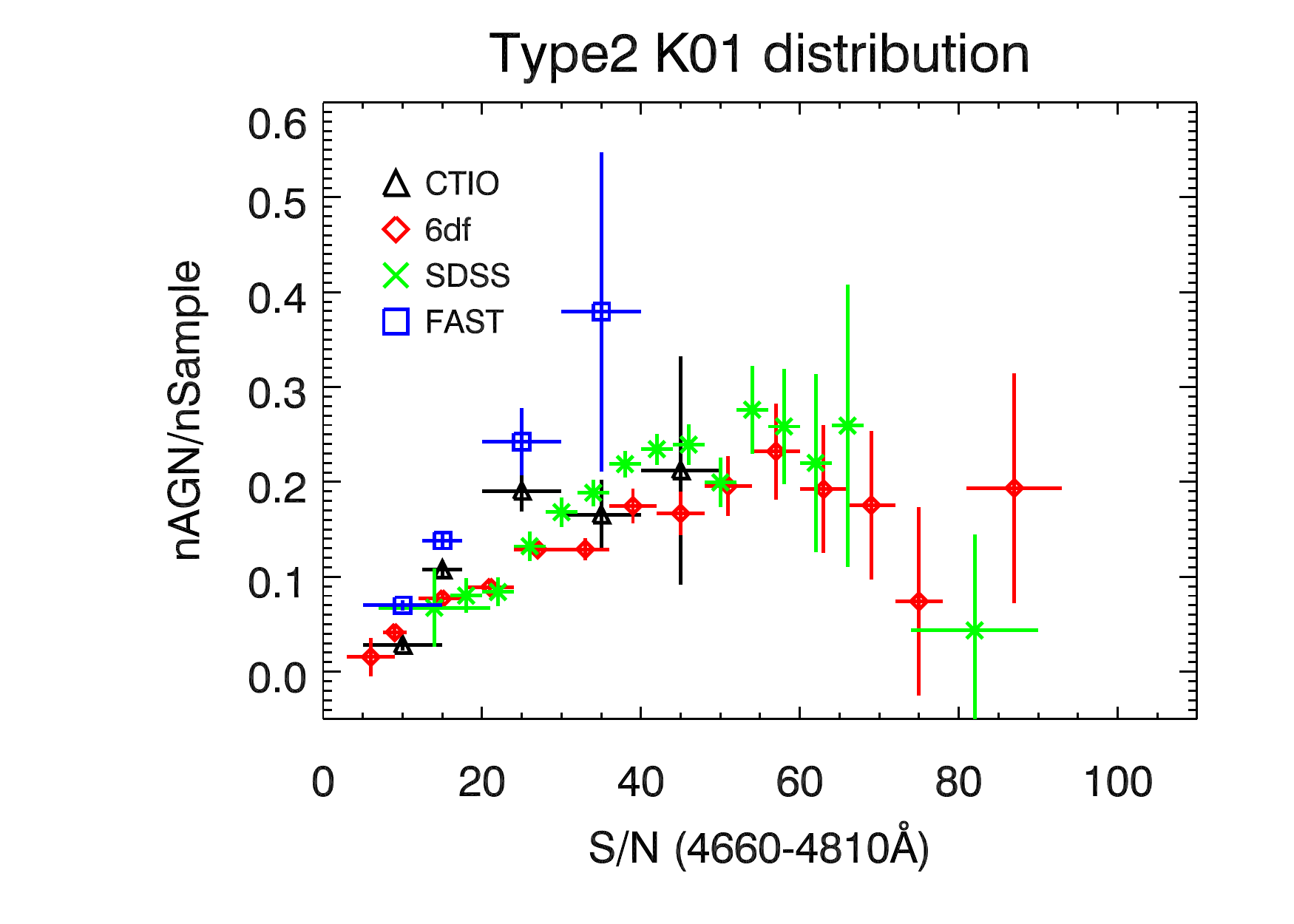}
\includegraphics[width=0.45\textwidth,angle=0]{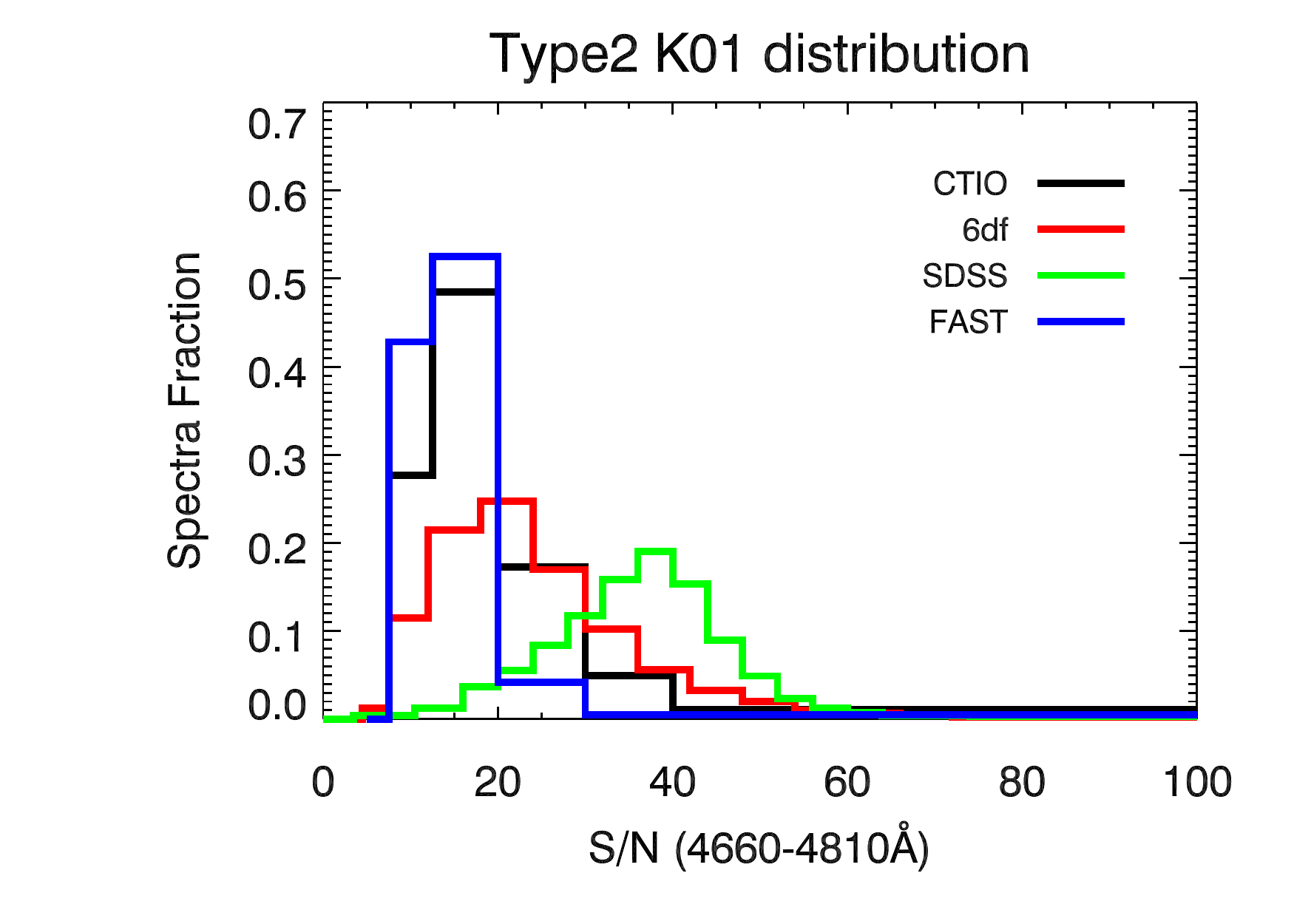}
\includegraphics[width=0.45\textwidth,angle=0]{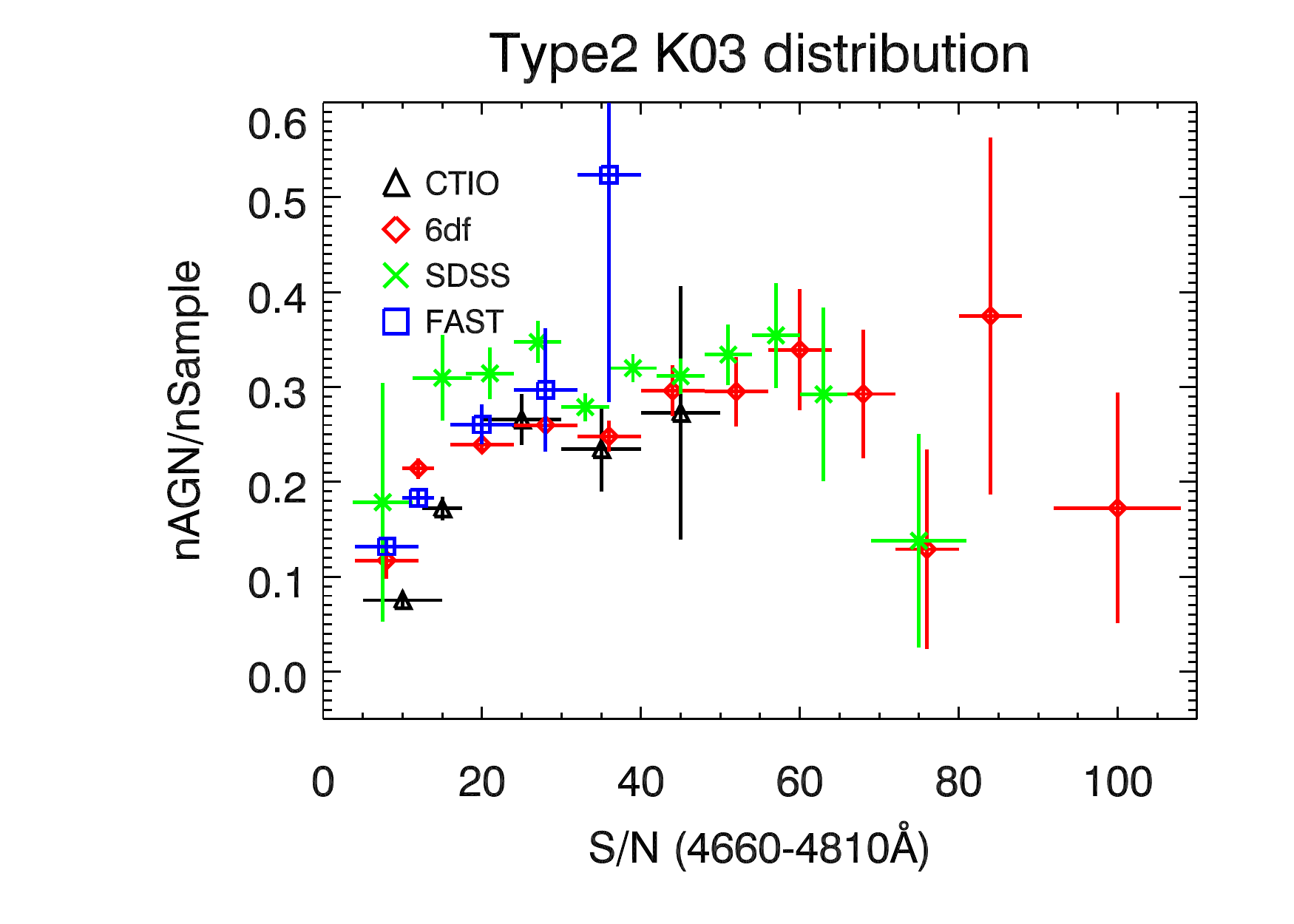}
\includegraphics[width=0.45\textwidth,angle=0]{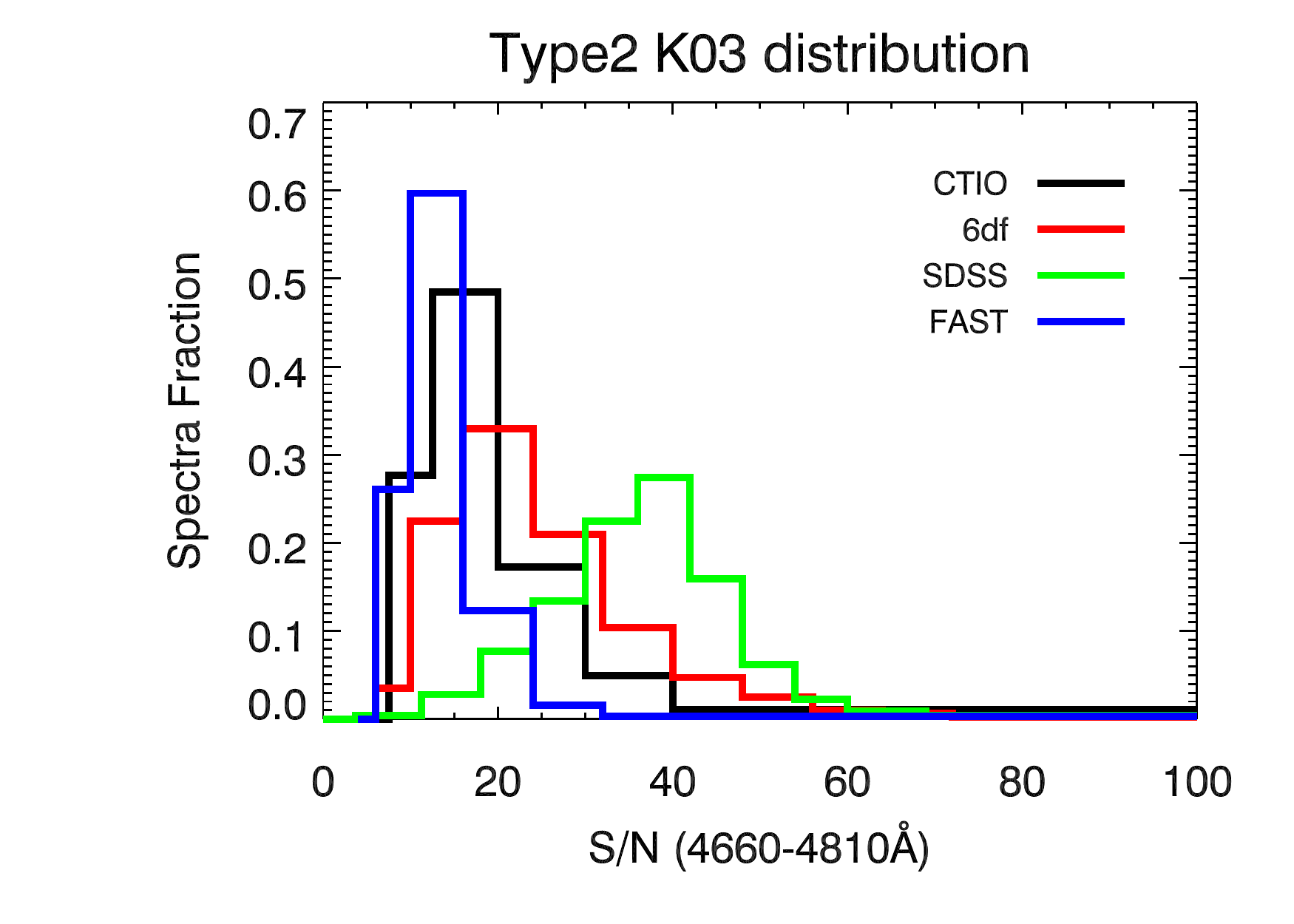}
\end{center}
\caption{The S/N of the spectra near the \Ha\ (for Type 1 AGNs) and \Hb\ (for Type 2 AGNs) lines for the AGNs in our catalog. The left plots show the AGN fraction as a function of S/N. The right plots show the histogram of the S/N distribution. The top pair are for Type 1 AGNs, the middle pair are for Type 2 AGNs satisfying \citet{Kewley01} criteria, and the bottom pair are for Type 2 AGNs satisfying \citet{Kauffmann03} criteria. The AGN fraction increases with S/N and flattens out at S/N $\gtrsim$ 50 for Type 2 AGNs and S/N $\gtrsim$ 75 for Type 1 AGNs; Beyond this value, the AGN fractions for the different subsamples are consistent with each other. SDSS has the highest S/N values, followed by 6dF. FAST and CTIO have comparable distributions of S/N and have very few spectra with S/N $\gtrsim$ 50.}
\label{fig:SN_AGN}
\end{figure}

Since AGN identification depends on emission line widths and ratios, the S/N of spectral lines affects the detection rates. Further analysis shows that these differences are due to overall, continuum, S/N of the spectrum. While the line S/N is determined by a mixture of spectral quality and the strength of the emission line in a spectrum, the continuum S/N does not depend on the line emission. The continuum S/N is a measure of the noise at a given wavelength, and the ability to detect a line at that wavelength. The left column of Figure~\ref{fig:SN_AGN} shows the AGN detection rate versus continuum S/N for, from top to bottom, Type 1 AGNs, Type 2 AGNs which meet the \citet{Kewley01} criteria, and Type 2 AGNs which meet the \citet{Kauffmann03} criteria. Since the continuum S/N is wavelength dependent, we choose to evaluate the continuum S/N between 6350 and 6540 \AA\ (near the \Ha\ line) for Type 1 AGNs. For Type 2 AGNs, we use evaluate the continuum S/N between 4660 and 4810 \AA, i.e. near the \Hb\ line, since \Hb\ is the weakest of the four lines used in Type 2 AGN identification.

The AGN detection fraction increases with S/N for all subsamples until a saturation point in the signal-noise-ratio, S/N (6350-6540 \AA) $\sim$50 for Type 2 AGNs and S/N (4660-48109\AA) $\sim$75 for Type 1 AGNs, at which point the AGN fraction flattens out. In addition, for S/N above the saturation value, 6dF and SDSS have consistent AGN detection fractions. FAST and CTIO subsamples do not have enough data to evaluate but their behavior is likely similar to 6dF. The similarity of the trend between the different subsamples for the different AGN types indicates that the overall spectral signal-to-noise is driving the AGN detection fraction. If the line S/N cut is changed, all the rates move up and down but the shape of the distribution remains the same. 

\begin{figure}[thb]
\includegraphics[scale=0.7]{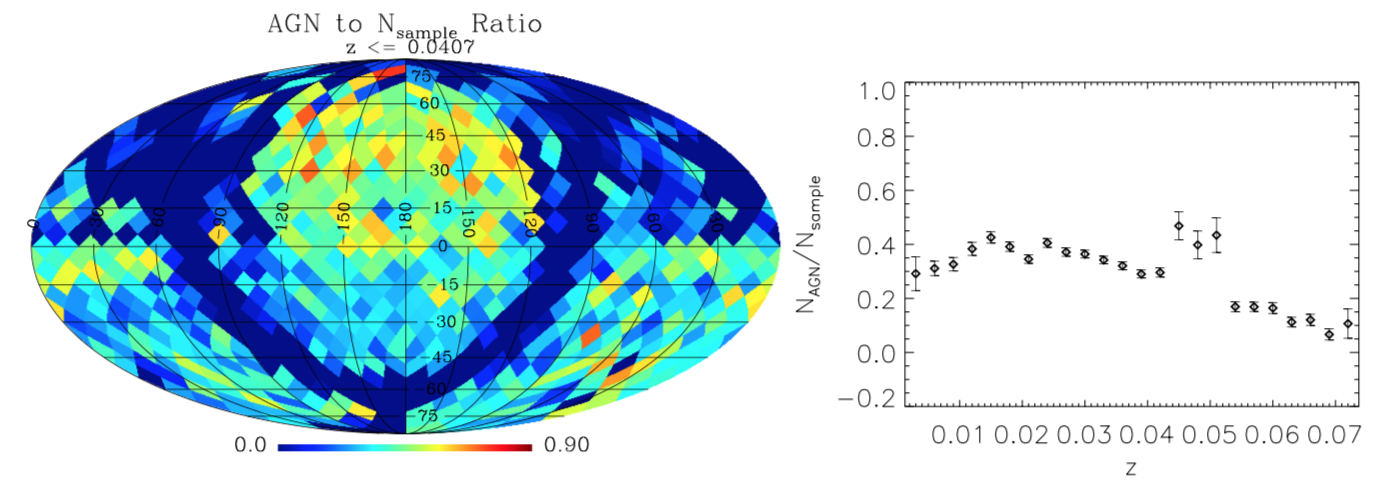}
\caption{Ratios of the number of identified AGNs to the number of galaxies in our spectroscopic sample, across the sky (left panel, the sky has been divided into 768 equal area regions and the color indicates the AGN fraction), and with respect to redshift (right panel). In these plots, we have combined the broad line AGNs with the narrow line AGNs which satisfy the \citet{Kauffmann03} criteria. The inhomogeneity across the sky is largely due to the differences in the AGN detection rates of the subsamples; the inhomogeneity in redshift is due to the telluric contamination of 6dF, FAST, and CTIO samples.}
\label{fig:AGNratio}
\end{figure}

\begin{figure}[htb]
\begin{center}
\includegraphics[scale=0.7]{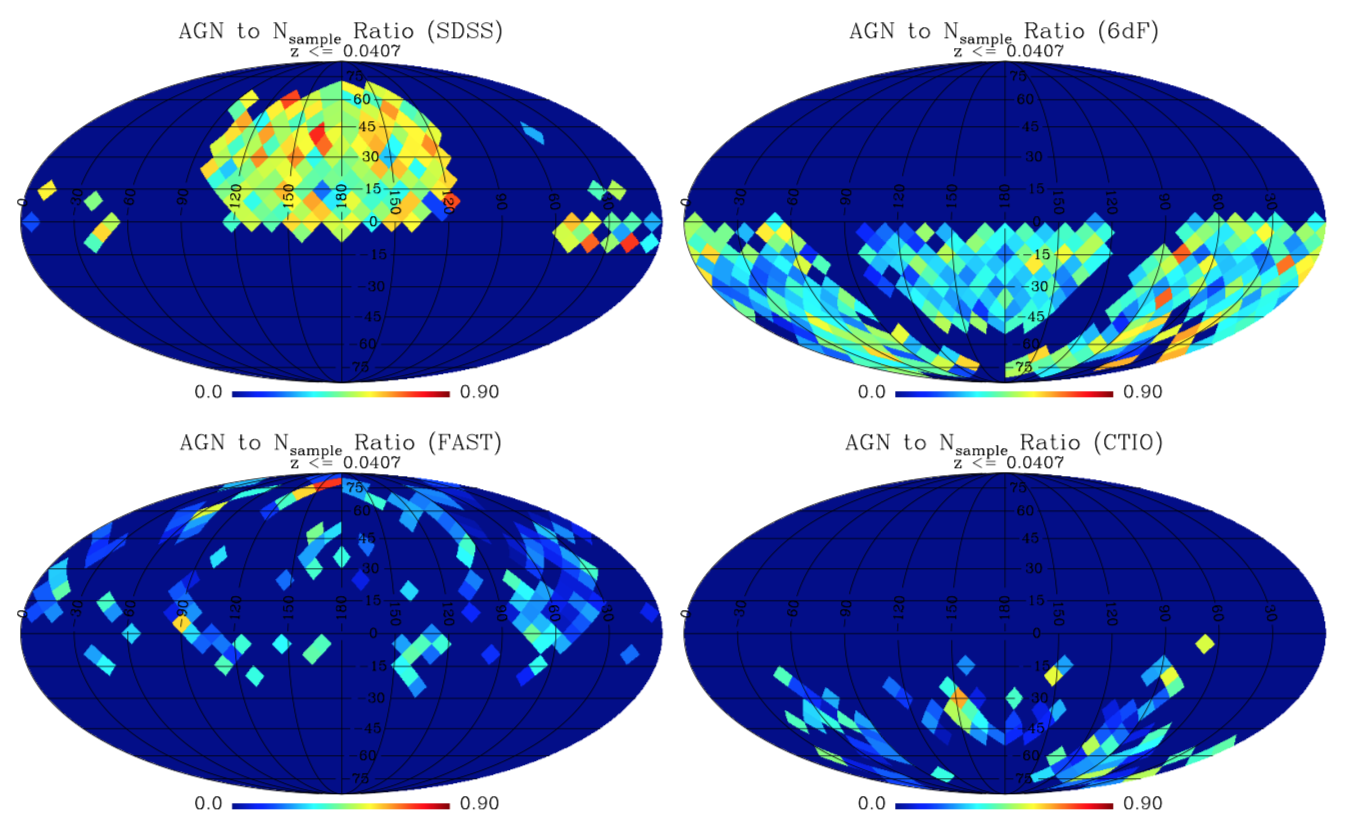}
\end{center}
\caption{The AGN fractions (the number of identified AGNs divided by the number of galaxies with spectra) across the sky (divided into 768 equal area regions), for each spectral subsample. Dark blue indicates pixels with fewer than 10 galaxies in 2MRS. In these plots, we have combined the broad line AGNs with the narrow line AGNs which satisfy the \citet{Kauffmann03} criteria. The AGN fractions are uniform for each subsample but the average AGN fraction differs between sample with SDSS having the highest and 6dF, FAST, and CTIO having lower fractions.}
\label{fig:AGNratio_subsamples_sky}
\end{figure}

\begin{figure}[htb]
\begin{center}
\includegraphics[scale=0.7]{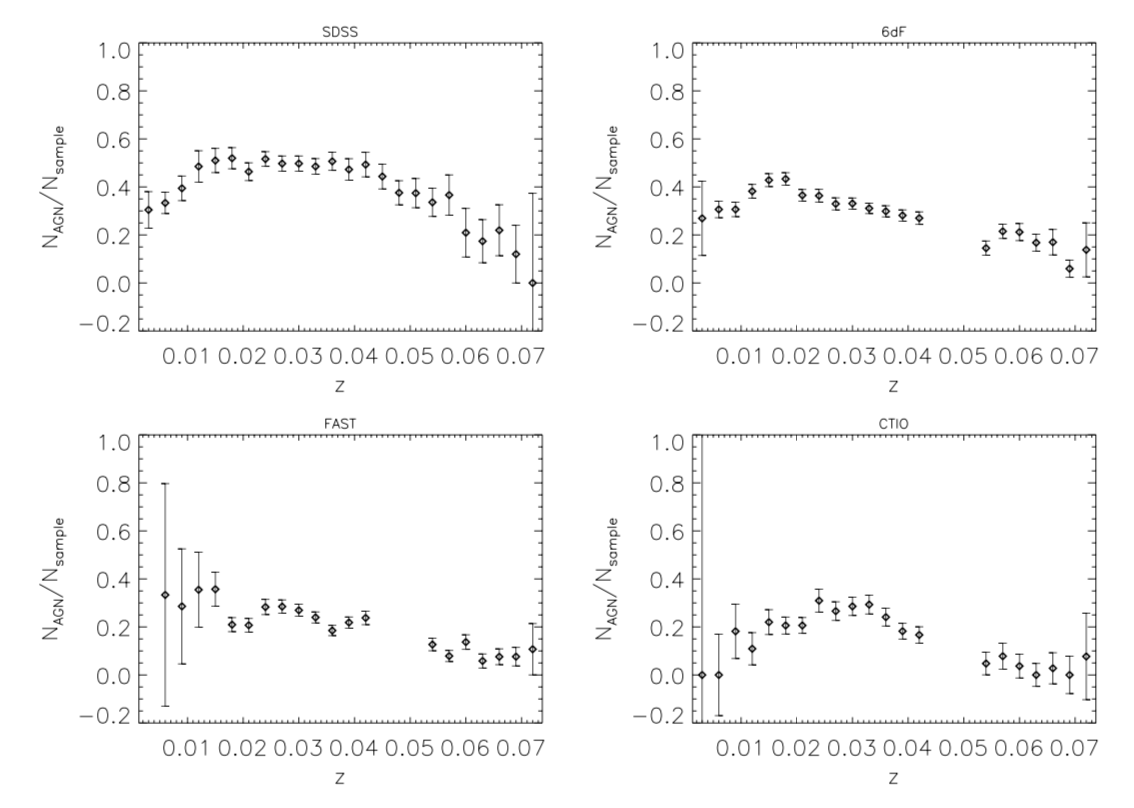}
\end{center}
\caption{The AGN fractions (the number of identified AGNs divided by the number of galaxies with spectra) versus redshift, for each spectral subsample. In these plots, we have combined the broad line AGNs with the narrow line AGNs which satisfy the \citet{Kauffmann03} criteria. The AGN fractions are uniform for each subsample but the average AGN fraction differs between sample with SDSS having the highest and 6dF, FAST, and CTIO having lower fractions. The 6dF, FAST, and CTIO samples show a gap in $z$ due to telluric contamination of the \NII-\Ha\ complex.}
\label{fig:AGNratio_subsamples_z}
\end{figure}

The right panels of Figure~\ref{fig:SN_AGN} show the distribution of continuum S/N for the four subsamples. As expected, the SDSS sample has the highest overall S/N, while 6dF, FAST, and CTIO show more similar distributions. Consequently, our catalog contains inhomogeneities due to the differences in sky coverage of the subsamples. Figure~\ref{fig:AGNratio} shows the AGN fractions (the number of identified AGN divided by the number of galaxies in our spectroscopic sample) for our catalog, across the sky (left panel) and in redshift (right panel). In each pixel on the sky and redshift bin, the number of AGNs is the sum of the number of broad line AGNs and the number of narrow line AGNs which satisfy the \citet{Kauffmann03} criteria. Assuming that  AGNs trace the large scale structure in which they are located, the AGNs-to-galaxies ratio should be uniform across the sky and in redshift. In the Southern hemisphere where there are fewer NED (and other missing) galaxies and 6dF dominates the available sample, we find that the AGN fraction is fairly uniform. The dark band traces the Galactic plane which we do not include our sample. In the North, the ratio is highest within the SDSS footprint, as expected given that SDSS spectra are the best among the subsamples, and is lower in other regions. The completeness with respect to redshift is fairly uniform except for the $ 0.0407< z < 0.0511$ region where the 6dF, FAST, and CTIO spectra are contaminated by the telluric line and we have only the SDSS sample, and beyond the range with the telluric contamination, where 2MRS itself becomes largely incomplete due to the requirement on $K_s$ magnitude. Although our AGN catalog has spatial and redshift inhomogeneities as a whole, we find that within each subsample, the AGN fraction is fairly uniform both spatially and in redshift, as shown in Figures~\ref{fig:AGNratio_subsamples_sky} and~\ref{fig:AGNratio_subsamples_z}.

\begin{figure}[thb]
\includegraphics[scale=0.7]{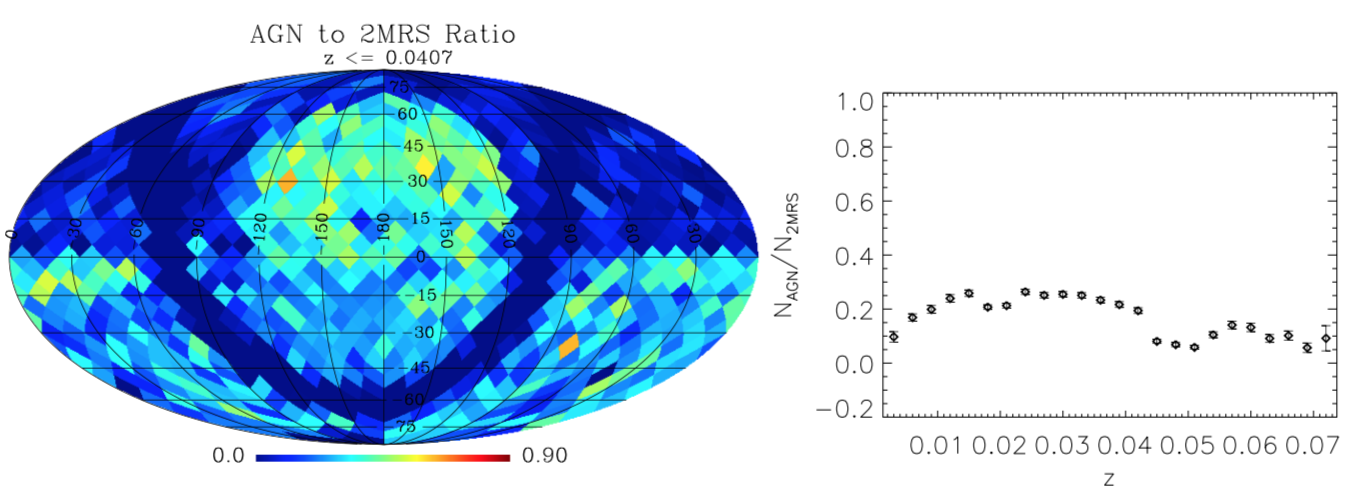}
\caption{Ratios of the number of identified AGNs to the number of galaxies in 2MRS, across the sky (left panel, the sky has been divided into 768 equal area regions and the color indicates the AGN fraction), and with respect to redshift (right panel). In these plots, we have combined the broad line AGNs with the narrow line AGNs which satisfy the \citet{Kauffmann03} criteria. The inhomogeneity across the sky is due the galaxies for which we do not have spectra (corrected for telluric contamination) and to the differences in the AGN detection rates of the subsamples, and that in redshift is due to the telluric contamination of 6dF, FAST, and CTIO samples.}
\label{fig:AGNratio_2MRS}
\end{figure}

The differences in AGN detection rate due to data quality affects not only catalogs which use inhomogeneous data but also analyses which compare results from different surveys. Thus, these effects must be taken into consideration when using any publicly available catalog, depending on the aims of the analysis. Our catalog contains additional inhomogeneity and incompleteness due to the missing or telluric contaminated spectra. These are shown in Figure~\ref{fig:AGNratio_2MRS}, where we plot the ratio of the number of identified AGNs to the number of galaxies in the parent 2MRS catalog, across the sky and in redshift. In the next section, we develop a method to statistically correct for the inhomogeneities and incompleteness due to the galaxies without spectra and those insufficient spectral signal-to-noise.

\subsection{A Statistical Correction for Incompleteness and Inhomogeneity Due to Spectral Quality}
\label{sec:statcat}

If this catalog is to be used in a statistical study, e.g. for correlations with the arrival direction of cosmic rays, the incompleteness and inhomogeneities must be accounted for to be able to fully understand the results. We hypothesize that the inhomogeneity and incompleteness (relative to the saturation rate) are primarily due to spectra with low continuum signal-to-noise (S/N). As seen in Figure~\ref{fig:SN_AGN}, the AGN fraction increases with S/N until a saturation value beyond which the fraction remains constant. This qualitative phenomenon is seen for Type 1 and Type 2 AGNs, with both the \citet{Kewley01} and \citet{Kauffmann03} criteria, for all the subsamples. Furthermore, after the saturation point, all the subsamples have AGN fractions consistent within error. We statistically recover the AGNs that would have been identified if their spectra had been of sufficient quality as follows:

We parametrize the AGN fraction as a function of the continuum S/N using a linear fit below the saturation S/N value, and a constant above, for Type 1 AGNs and Type 2 AGNs satisfying the \citet{Kewley01} criteria.\footnote{We use the \citet{Kewley01} criteria in this analysis because they yield a purer sample of galaxies with AGN activity. Furthermore, the AGN fraction for Type 2 AGNs satisfying the \citet{Kauffmann03} criteria seem to follow a curve rather than a line and requires a more complicated analysis.} In doing so, we obtain the following relations:
\begin{align*}
R_{fT1} = & (SN1)*0.002412-0.01400&  \; {\rm for} \, SN1 < 76 \\
             = & 0.1693 \equiv R_{sT1}&  \; {\rm for} \, SN1 \geq 76 \\
R_{fT1} =  & (SN2)*0.004621-0.01214 & \; {\rm for} \, SN2 < 50 \\
             = & 0.2432 \equiv R_{sT2}&  \; {\rm for} \, SN2 \geq 50 \\
\end{align*}
where $SN1$ ($SN2$) is the continuum signal-to-noise for wavelengths 6350-6540 \AA\ (4660-4810 \AA), i.e. near the \Ha\ (\Hb) line, $R_{fT1}$ ($R_{fT2}$) is the AGN fraction for the given galaxy's S/N value for Type 1 (Type 2) AGNs, and $R_{sT1}$ ($R_{sT2}$) is the saturation value for the Type 1 (Type 2) AGN fraction. If a non-AGN has a S/N (6350-6540 \AA) value above 76, we assign it a Type 1 AGN likelihood of zero. Similarly, if it has a S/N (4660-4810 \AA) above 50, we assign a Type 2 AGN likelihood of zero. If a galaxy's S/N values near the \Ha\ and \Hb\ lines are less than the saturation values, we assign likelihoods as follows:
\begin{align*}
&L_{T1}  = \frac{R_{sT1}-R_{fT1}}{1-R_{fT1}-R_{fT2}} & \\
&L_{T2}  = \frac{R_{sT2}-R_{fT2}}{1-R_{fT1}-R_{fT2}}, &
\end{align*}
where $L_{sT1}$ and $L_{sT2}$ are the likelihoods for the galaxy to be a Type 1 and Type 2 AGN, respectively. For galaxies which are not in our spectroscopic sample, we assign the saturation AGN fractions as their AGN likelihoods. The probability for a galaxy to be an AGN is the sum of the likelihoods for Type 1 and Type 2 AGNs. The galaxies identified as AGNs are assigned an AGN likelihood of 1.0. Examples of the likelihood assignments for the non-AGNs and galaxies with missing spectra are given in Tables~\ref{tab:nonAGNs} and \ref{tab:missing}, respectively. These likelihoods are intended to statistically correct for both the inhomogeneity and incompleteness. If different criteria are used for identifying AGNs, e.g. different line S/N, this analysis should be repeated to assess the incompleteness and inhomogeneity.

\begin{figure}[thb]
\includegraphics[scale=0.7]{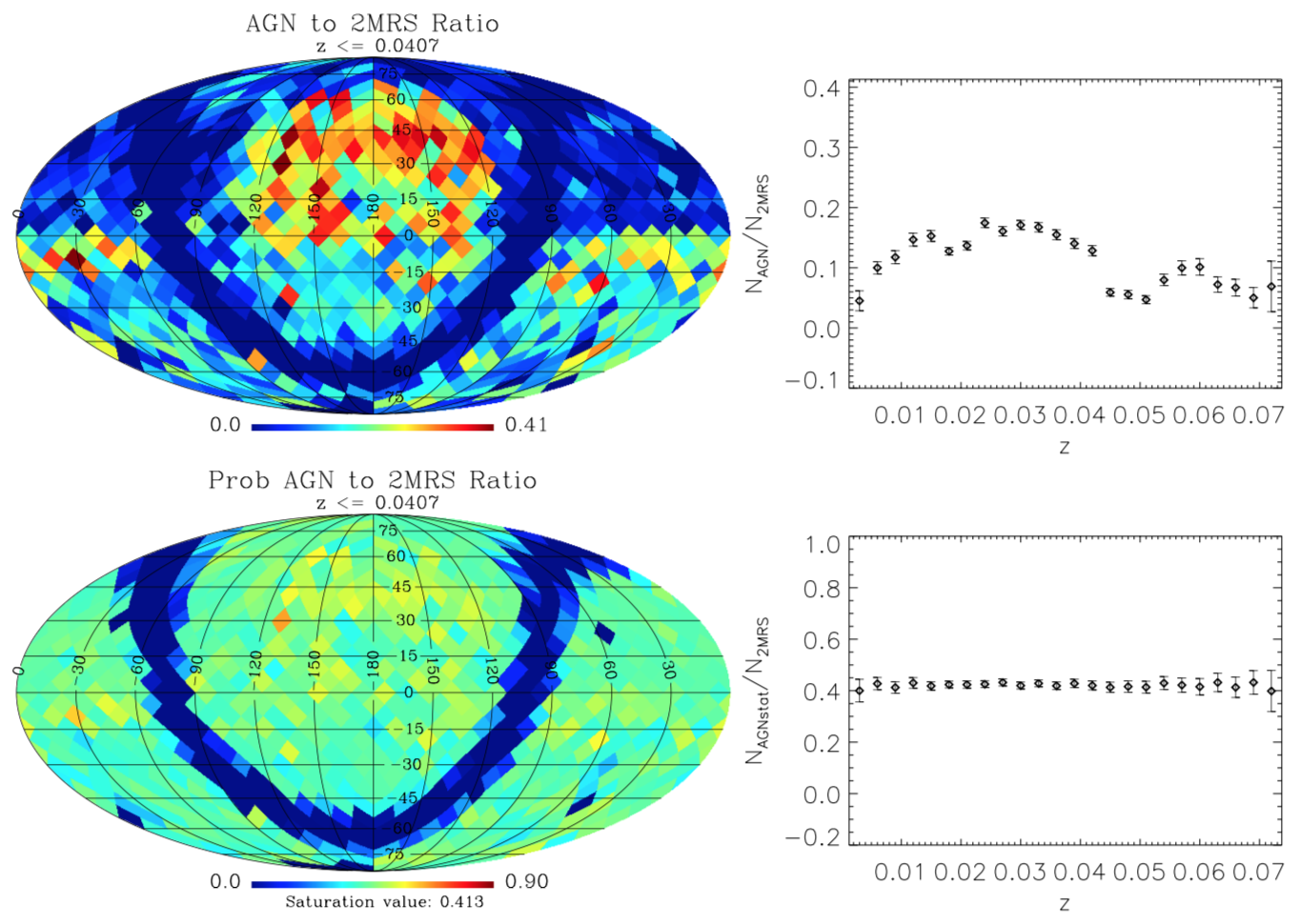}
\caption{Ratios of the number of AGNs to the number of galaxies in 2MRS, across the sky (left panels, the sky has been divided into 768 equal area regions and the color indicates the AGN fraction), and with respect to redshift (right panels). In these plots, we have combined the broad line AGNs with the narrow line AGNs which satisfy the \citet{Kewley01} criteria. The top plots show the AGN fractions with the AGNs satisfying the emission line criteria. We have set the AGN saturation value as the maximum for the plotted ranges. The bottom plots show the AGN fraction where the number of AGNs is defined as the sum of the AGN likelihoods. The statistically enhanced catalog has a homogeneous AGN fraction across the sky and in redshift, as expected assuming that AGNs trace large scale structure. The ratios also fall around the saturation value, now in the middle of the plotted range, indicating that the incompleteness has also been statistically accounted for.}
\label{fig:AGNratio_2MRS_K01_cor}
\end{figure}

Figure~\ref{fig:AGNratio_2MRS_K01_cor} shows the number of AGNs (Type 1 and Type 2 satisfying the \citet{Kewley01} criteria) divided by the number of galaxies in 2MRS, i.e. the AGN fraction, across the sky (left) and with respect to redshift (right). In the top plots, the number of AGNs is simply the number which pass the emission line criteria. These identified AGNs are inhomogeneous across the sky and in redshift. In addition, the ratios across the sky and in redshift are below the saturation value, set to be the maximum of the range. In the bottom plots, the number of AGNs is defined as the sum of the AGN likelihoods. The ratio of these ``statistical" AGNs to the number of galaxies in 2MRS is uniform across both the sky and in redshift. Additionally, the ratios fall around the saturation value, now in the middle of the plotted range. Therefore, the AGN likelihoods statistically compensate for the inhomogeneity and incompleteness of the identified AGNs due to spectral quality and can be used in statistical studies. The analysis also identifies galaxies whose AGN status needs to be clarified with higher quality spectroscopy in order to have a complete and homogeneous AGN catalog.

This result validates the assumption that the inhomogeneity and incompleteness are dominated by spectral quality. Furthermore, these results indicate that, in order for an optical catalog to be complete, the spectra should have S/N (6350-6540 \AA) $\gtrsim$ 76 and S/N (4660-4810 \AA) $\gtrsim$ 50. The higher S/N requirement for Type 1 is consistent with the fact that the broad lines typically have lower peaks than narrow lines, and are harder to distinguish from the noise. 

\begin{table}
 \caption{Non-AGN Galaxies (Illustrative Extract)}
 \label{tab:nonAGNs}
\footnotesize
\begin{center}
\begin{tabular}{|l|r|r|r|l|l|r|r|r|r|r|}
\hline
  \multicolumn{1}{|c|}{TMID} &
  \multicolumn{1}{c|}{RA (deg)} &
  \multicolumn{1}{c|}{DEC (deg)} &
  \multicolumn{1}{c|}{V (km/s)} &
  \multicolumn{1}{c|}{CAT} &
  \multicolumn{1}{c|}{TYPE} &
  \multicolumn{1}{c|}{S/N} &
  \multicolumn{1}{c|}{S/N} & 
  \multicolumn{1}{c|}{PROBT1} &
  \multicolumn{1}{c|}{PROBT2} &
  \multicolumn{1}{c|}{PROBAGN} \\
 \multicolumn{1}{|c|}{} &
 \multicolumn{1}{c|}{} &
 \multicolumn{1}{c|}{} &
 \multicolumn{1}{c|}{} &
 \multicolumn{1}{c|}{} &
 \multicolumn{1}{c|}{} &
 \multicolumn{1}{c|}{(4660-4810\AA)} &
 \multicolumn{1}{c|}{ (6350-6540\AA)} &
 \multicolumn{1}{c|}{} &
 \multicolumn{1}{c|}{} &
 \multicolumn{1}{c|}{} \\
\hline
  00000865-0622263 & 0.03601 & -6.37399 & 6531 & 6 & Emi & 16.78 & 27.80 &0.1356 & 0.1791 & 0.3147\\
  00002363-4701076 & 0.09836 & -47.01881 & 5991 & 6 & -- & 24.10 & 53.66 &0.0708 & 0.1573 & 0.2281\\
  00005858-3336429 & 0.24409 & -33.61195 & 6914 & 6 & Emi & 14.83 & 19.13 & 0.1546 & 0.1832 & 0.3378\\
  00010289-4319496 & 0.26204 & -43.33044 & 11627 & 6 & -- & 22.78 & 38.48& 0.1126 & 0.1565 & 0.2691\\
  00010597-5359303 & 0.27494 & -53.99181 & 9423 & C & Emi & 20.32 & 28.78 & 0.1358 & 0.1636 & 0.2994\\
  00011748-5300348 & 0.32276 & -53.00967 & 9724 & 6 & Emi & 12.91 & 18.80 & 0.1538 & 0.1911 & 0.3450\\
  00013295-0726099 & 0.38729 & -7.43614 & 8836 & 6 & Emi & 26.41 & 40.53 & 0.1094 & 0.1394 & 0.2488\\
  00014077+5344492 & 0.41976 & 53.747 & 10813 & F & -- & 15.46 & 24.71 & 0.1421 & 0.1833 & 0.3254\\
  00015102-6911394 & 0.46241 & -69.19435 & 18236 & 6 & -- & 17.38 & 20.91 & 0.1525 & 0.1730 & 0.3256\\
  00020544-3037082 & 0.52273 & -30.61901 & 9063 & 6 & -- & 20.02 & 30.24 & 0.1320 & 0.1657 & 0.2976\\
  00024370+2725454 & 0.68219 & 27.42926 & 7547 & F & -- & 9.46 & 14.26 & 0.1612 & 0.2028 & 0.3640\\
  00024956+0424345 & 0.70653 & 4.40957 & 11778 & F & -- & 14.70 & 20.26 & 0.1519 & 0.1843 & 0.3362\\
  00030496-4210429 & 0.77077 & -42.17865 & 11816 & 6 & Emi & 21.67 & 22.68 & 0.1518 & 0.1546 & 0.3064\\
  00031331+5352149 & 0.80545 & 53.87077 & 11772 & F & -- & 5.85 & 9.56 & 0.1684 & 0.2144 & 0.3827\\
  00032138-5004494 & 0.83896 & -50.08049 & 10333 & 6 & Emi & 38.00 & 55.42 & 0.0717 & 0.0801 & 0.1517\\
  00032452-4947304 & 0.85217 & -49.79181 & 11258 & 6 & Emi & 34.34 & 61.39 & 0.0507 & 0.1041 & 0.1548\\
  00033726+6919064 & 0.905 & 69.31848 & 6864 & F & -- & 1.31 & 7.89 & 0.1682 & 0.2304 & 0.3985\\
  00044222-3029007 & 1.17596 & -30.48364 & 7994 & 6 & -- & 61.49 & 79.54 & 0.0 & 0.0 & 0.0\\
  00500091-1440566 & 12.50379 & -14.68239 & 15997 & 6 & Emi & 44.61 & 76.22 & 0.0 & 0.0376 & 0.0376\\
  03175957+4131127 & 49.49819 & 41.52027 & 3315 & S & -- & 55.00 & 73.20 & 0.0114 & 0.0 & 0.0114\\  
\hline\end{tabular}
\end{center}
  \begin{tablenotes}
      \small
      \item TMID: Two MASS ID; RA: right ascension; DEC: declination; V: velocity; CAT: Spectral source, 6=6dF, S=SDSS, F=FAST, C=CTIO; Type: Galaxy type, Emi = Emission line galaxy, -- = Non-emission line galaxy; S/N (4660-4810\AA): Signal-to-noise ratio of the spectrum near the \Hb\ line; S/N (6350-6540\AA): Signal-to-noise ratio of the spectrum near the \Ha\ line;  PROBT1: Likelihood to be a Type 1 AGN;  PROBT2: Likelihood to be a Type 2 AGN; PROBAGN: Likelihood to be an AGN.  Table 4 is published in its entirety in the electronic edition of the {\it Astrophysical Journal}.  A portion is shown here for guidance regarding its form and content. The full table also includes the emission line fluxes and errors of all the non-AGN galaxies.
      \end{tablenotes}  
\end{table}

\begin{table}
 \caption{The 2MRS Galaxies Not in Our Spectroscopic Sample (Illustrative Extract)}
 \label{tab:missing}
\footnotesize
\begin{center}
\begin{tabular}{|l|r|r|r|r|l|r|r|r|}
\hline
  \multicolumn{1}{|c|}{TMID} &
  \multicolumn{1}{c|}{RA (deg)} &
  \multicolumn{1}{c|}{DEC (deg)} &
  \multicolumn{1}{c|}{$M_K$} &
  \multicolumn{1}{c|}{V (km/s)} &
  \multicolumn{1}{c|}{CAT} &
  \multicolumn{1}{c|}{PROBT1} &
  \multicolumn{1}{c|}{PROBT2} &
  \multicolumn{1}{c|}{PROBAGN} \\
\hline
  01100566-4754104 & 17.52361 & -47.90289 & 11.754 & 16579 & N & 0.1693 & 0.2432 & 0.4125 \\
  01550669+2011390 & 28.77795 & 20.19413 & 11.754 & 8994 & N & 0.1693 & 0.2432 & 0.4125 \\
  02344425+3429532 & 38.68447 & 34.49808 & 11.754 & 7735 & N & 0.1693 & 0.2432 &0.4125 \\
  05440707-4838401 & 86.02939 & -48.64445 & 11.754 & 14428 & 6 & 0.1693 & 0.2432 & 0.4125 \\
  07085955-3337448 & 107.24809 & -33.62906 & 11.754 & 13248 & N & 0.1693 & 0.2432 & 0.4125 \\
  07195464+5534213 & 109.97761 & 55.57257 & 11.754 & 12714 & N & 0.1693 & 0.2432 & 0.4125\\
  07431408+5458078 & 115.80865 & 54.96876 & 11.754 & 10289 & N & 0.1693 & 0.2432 & 0.4125\\
  10122225-1631129 & 153.09288 & -16.52024 & 11.754 & 14836 & 6 & 0.1693 & 0.2432 & 0.4125 \\
  10353083-2722496 & 158.87851 & -27.38052 & 11.754 & 4425 & N & 0.1693 & 0.2432 & 0.4125 \\
  11320692+7048556 & 173.02899 & 70.81548 & 11.754 & 16070 & N & 0.1693 & 0.2432 & 0.4125 \\
  19024004-1715194 & 285.66675 & -17.25544 & 11.754 & 13913 & 6 & 0.1693 & 0.2432 & 0.4125\\
  21002390+0927135 & 315.09958 & 9.45379 & 11.754 & 9033 & O & 0.1693 & 0.2432 & 0.4125\\
  21472088-1732573 & 326.837 & -17.54931 & 11.753 & 10373 & N & 0.1693 & 0.2432 & 0.4125\\
  22031497-5558516 & 330.81238 & -55.98096 & 11.753 & 6865 & N & 0.1693 & 0.2432 & 0.4125\\
  22085077-3507030 & 332.21158 & -35.11749 & 11.753 & 9617 & N & 0.1693 & 0.2432 & 0.4125\\
  22365419-2656461 & 339.22577 & -26.9461 & 11.753 & 12766 & 6 & 0.1693 & 0.2432 & 0.4125\\
  23202242+5138450 & 350.0934 & 51.64578 & 11.753 & 13689 & F & 0.1693 & 0.2432 & 0.4125\\
  23270825-4936242 & 351.78433 & -49.6067 & 11.754 & 26950 & N & 0.1693 & 0.2432 & 0.4125\\
  23475980-1558381 & 356.99915 & -15.97725 & 11.753 & 15647 & 6 & 0.1693 & 0.2432 & 0.4125\\
  23520287+2102068 & 358.01202 & 21.03524 & 11.754 & 13233 & F & 0.1693 & 0.2432 & 0.4125\\
\hline\end{tabular}
\end{center}
  \begin{tablenotes}
      \small
      \item TMID: Two MASS ID; RA: right ascension; DEC: declination; $M_K$: Extinction corrected $K_s$ isophotal magnitude; V: velocity; CAT: Source of 2MRS redshift; PROBT1: Likelihood to be a Type 1 AGN;  PROBT2: Likelihood to be a Type 2 AGN; PROBAGN: Likelihood to be an AGN.  \end{tablenotes}  
\end{table}

\section{Conclusions}
\label{sec:conclusions}
We have constructed an all-sky catalog of nearby AGNs, uniformly selected from the parent sample of 2MRS galaxies and classified using optical spectra. The catalog consists of 1929 broad line AGNs and 6562 narrow line AGNs which satisfy the \citet{Kauffmann03} criteria, of which 3607 also satisfy the \citet{Kewley01}. For each AGN and emission line galaxy, we report the line widths, fluxes, flux errors, and signal-to-noise ratios fo the AGN identification lines, enabling the user to customize his/her own AGN catalog. We also provide an assessment of the completeness and homogeneity of our catalog across the sky and in redshift. Although we start with a homogeneous, complete parent sample and we process the spectra uniformly, inhomogeneity remains because we have different subsamples whose spectra were taken with different instruments, which have different spectral resolution and signal-to-noise. The differences in data quality affect not only the overall AGN detection rate but also the broad line to narrow line AGN ratio. We find that a spectrum should have a continuum signal-to-noise $\gtrsim$75 ($\gtrsim$50) near the lines of interest in order to have full efficiency for detecting Type 1 (Type 2) AGNs.

If an analysis needs individual AGNs, e.g. for extragalactic water maser searches, this catalog can be used as is or customized with different selection criteria using the spectroscopic measurements provided. The \citet{Kauffmann03} criteria selects the galaxies with any AGN contribution while the \citet{Kewley01} criteria selects those with emission lines dominated by emission from AGN activity. Increasing the signal-to-noise requirement will improve the purity of the sample at the expense of the completeness and homogeneity. If the catalog is to be used in a statistical study, e.g. for correlations with the arrival directions of ultra-high energy cosmic rays, the incompleteness and inhomogeneity must be taken into account.

In order to make this catalog better suited for statistical studies, we have assigned Type 1, Type 2, and total AGN probabilities for the non-AGNs in our spectroscopic sample and for the 2MRS galaxies for which we do not have telluric corrected spectra. After these corrections, the AGN fractions are uniform across the sky and in redshift. This catalog is thus suitable for statistical studies.  The analysis also paves the way for a truly complete and homogeneous nearby AGN catalog by identifying galaxies whose AGN status needs to be verified with higher quality spectra, quantifying the spectral quality necessary to do so. This work also underscores the importance of accounting for differences in spectral quality and resolution before comparing the AGN detection fractions and Type 1 to Type 2 AGN ratios and from different optical AGN surveys and when conducting populations studies of AGNs from available catalogs.

\acknowledgments

We thank the anonymous referee for his/her helpful comments. We are grateful to Lucas Macri for providing us with, and for helping us better understand, the FAST and CTIO spectra, and to Heath Jones for help with the 6dF spectra. We thank Yuxiao Dai for checking our catalog and pointing us to salient information in the literature, Helen Treiber and Yanfei Zhang for their work with the NED galaxies, and John Huchra, Lincoln Greenhill, Jessica Mink, and Avanti Tilak for their involvement at the beginning of this project. We thank Jong-Hak Woo and Michael Blanton for discussions on AGN identification and SDSS spectral analysis, respectively. We also thank the anonymous referee for his or her suggestions. The research of GRF was supported by National Science Foundation grants NSF-PHY-1212538 and NSF-AST-1517319, and by the James Simons Foundation. 

Funding for SDSS-III has been provided by the Alfred P. Sloan Foundation, the Participating Institutions, the National Science Foundation, and the U.S. Department of Energy Office of Science. The SDSS-III web site is http://www.sdss3.org/.

SDSS-III is managed by the Astrophysical Research Consortium for the Participating Institutions of the SDSS-III Collaboration including the University of Arizona, the Brazilian Participation Group, Brookhaven National Laboratory, Carnegie Mellon University, University of Florida, the French Participation Group, the German Participation Group, Harvard University, the Instituto de Astrofisica de Canarias, the Michigan State/Notre Dame/JINA Participation Group, Johns Hopkins University, Lawrence Berkeley National Laboratory, Max Planck Institute for Astrophysics, Max Planck Institute for Extraterrestrial Physics, New Mexico State University, New York University, Ohio State University, Pennsylvania State University, University of Portsmouth, Princeton University, the Spanish Participation Group, University of Tokyo, University of Utah, Vanderbilt University, University of Virginia, University of Washington, and Yale University.



\end{document}